\documentclass{article}

\usepackage[english]{babel}
\usepackage[utf8]{inputenc}
\usepackage[T1]{fontenc}

\usepackage[letterpaper,top=3cm,bottom=2cm,left=3cm,right=3cm,marginparwidth=1.75cm]{geometry}

\usepackage{amsmath}
\usepackage{graphicx}
\usepackage[page,toc,titletoc,title]{appendix}
\usepackage[colorlinks=true, allcolors=blue]{hyperref}
\usepackage{stackengine}
\usepackage{physics}
\usepackage{amssymb} 
\usepackage{mathrsfs}
\usepackage{amsfonts}
\usepackage{amssymb}
\usepackage[useregional]{datetime2}
\usepackage[left]{lineno}
\usepackage{longtable}
\usepackage{booktabs}
\usepackage[framemethod=TikZ]{mdframed}
\usepackage[breakable]{tcolorbox}
\usepackage{csquotes}
\usepackage{pdflscape}
\usepackage{tabularx}
\usepackage{longtable}
\usepackage{makecell}
\usepackage{adjustbox}
\usepackage{array}
\usepackage{csquotes}
\usepackage{blindtext}
\usepackage[affil-it]{authblk}
\usepackage{setspace}
\usepackage{mathtools} 
\usepackage{accents}

\newcommand{\myvect}[1]{\accentset{\rightharpoonup}{#1}}

\usepackage[sort,semicolon,authoryear,round]{natbib}
\newcommand{\posscite}[1]{\citeauthor{#1}'s \citeyear{#1}}

\newcommand{\Cov}[3]{\mathrm{Cov}_{#1} \negmedspace \left( #2, #3 \right)}
\newcommand{\Var}[2]{\mathrm{Var}_{#1} \negmedspace \left( #2\right)}
\renewcommand{\eqref}[1]{Eq.\ref{#1}}
\newcommand{\noop}[1]{}

\newmdenv[
  topline=false,
  bottomline=false,
  skipabove=\topsep,
  skipbelow=\topsep,
  leftmargin=-10pt,
  rightmargin=-10pt,
  innertopmargin=0pt,
  innerbottommargin=0pt
]{siderules}

\title{The storage effect is not about bet-hedging or population stage-structure}
\date{}
\author[1,2,*]{Evan C. Johnson}
\author[3]{Oscar Godoy}
\author[1]{Alan Hastings}

\affil[1,]{Department of Environmental Science and Policy; University of California Davis; Davis, California 95616 USA}
\affil[2,]{Center for Population Biology; University of California Davis; Davis, California 95616 USA}
\affil[3,]{Department of Biology; Universidad de Cádiz; Puerto Real, 11510 Spain}
\affil[*]{Corresponding author: Evan Johnson, evcjohnson@ucdavis.edu}

\begin{document}
\maketitle
\clearpage

\section*{Abstract}

The storage effect is a well-known explanation for coexistence in temporally varying environments. Like many complex ecological theories, the storage effect is often used as an explanation for observed coexistence on the basis of heuristic understanding, rather than careful application of a detailed model. One interpretation states that species coexist by specializing on specific environmental states, and therefore must have a robust life-stage (e.g., long-lived adults, a seed-bank) in order to "wait it out" for favorable conditions. Here we show that this widely employed interpretation can be misleading. Multiple models show that stage-structure, long lifespans, and overlapping generations are neither necessary nor sufficient for the storage effect. In models where a robust life-stage does engender a storage effect, it does not do so by preventing stochastic extinction or by improving relative bet-hedging. A robust life-stage is best understood as one of many ways to fulfill an abstract condition for the storage effect: an interaction effect of environment and competition on per capita growth rates. Using a dataset of annual plants from a Mediterranean grassland in Spain, we show that such interaction effects occur between water availability and the number of germinant competitors, leading to storage in the absence of a persistent seed bank. Empiricists hoping to uncover the storage effect should look for interaction effects between environmental conditions and competition --- easily identifiable with multiple regression --- at all stages of a species' life-cycle.

\newpage

\tableofcontents

\newpage

\section{Introduction}
\label{Introduction}

The temporal storage effect (often simply called the storage effect) is a general explanation for how species can coexist by specializing on different states of a temporally fluctuating environment. The storage effect has an impressive resume. First, it formalized the concept of environmental niche partitioning, which has long been thought (e.g., \citealp{grinell1917niche}) to promote coexistence. Second, by showing that many species can coexist on just a single resource (\citealp{Chesson1994}, Eq. 91;  \citealp{miller2017evolutionary}), the storage effect provided a potential resolution to \posscite{hutchinson1961paradox} paradox of the plankton. Subsequent empirical investigations revealed that the storage effect is stabilizing and destabilizing in real communities (e.g. \citealp{angert2009functional}; \citealp{ellner2016quantify}; \citealp{zepeda2019fluctuation}). Third, the storage effect revealed the beneficent side of temporal environmental variation, which historically was thought to undermine persistence and coexistence (\citealp{Lewontin1969}; \citealp{may1974theory}). A number of recent papers have synthesized the simultaneously stabilizing and destabilizing effects of environmental variation (\citealp{adler2008environmental}; \citealp{schreiber2019rarity}; \citealp{pande2020mean}, \citealp{dean2020stochasticity}).

But what exactly \textit{is} the storage effect? The simple answer, "coexistence due to environmental niche differences" is a good start, but it is not sufficient ---  environmental niche differences alone do not promote coexistence (\citealp{chesson1997roles}). A common interpretation of the storage effect involves a robust life stage that can persist until the environmental becomes favorable, but as we will show, this is misleading.
Our contention in this paper is that the storage effect is inherently complicated: much like a p-value, there is no interpretation that is simultaneously concise, correct, and intuitive. Although the literature often speaks of the storage effect as a specific mechanism, it is a general phenomenon in actuality, such that a complete description of the storage effect requires abstractions like covariances, invader--resident comparisons, and interaction effects.

% It is not sufficient to point at the mathematical definition of the storage effect --- symbols are only meaningful once interpreted. 

% This is partially because the storage effect is a general phenomenon, such that a general description of the storage effect requires abstractions like covariances, invader--resident comparisons, and interaction effects.

The storage effect was discovered by \citet{chesson1981environmentalST} and coined by \citet{chesson1983coexistence}, though an analogous phenomenon had been previously noted in the context of population genetics (\citealp{Dempster1955}; \citealp{Haldane1963}; \citealp{gillespie1977natural}). In the 1980's, a sequence of papers (\citealp{chesson1981environmentalST}; \citealp{chesson1982stabilizing}; \citealp{chesson1983coexistence}; \citealp{chesson1984storage}; \citealp{warner1985coexistence}) analyzed a model of reef fish dynamics (\textit{the lottery model}) and repeatedly highlighted an interesting result: coexistence is not possible if both species have non-overlapping generations (i.e., if the adult fish survival probability equals zero). An analogous result was found for a model of annual plants (\citealp{Chesson1994}, Section 5): coexistence is not possible if neither species has a seed bank.

The seminal models of the storage effect --- the lottery model and annual plant model --- along with excerpts from the literature (Appendix \ref{The conventional interpretation: Excerpts from the literature}) may give the impression that the storage effect requires a robust life-stage that can "wait it out" for a good year; "Storage" refers to the vitality of a robust life-stage. However, one can construct models where the storage effect promotes coexistence, despite the absence of stage-structure and overlapping generations (e.g., \citealp{abrams1984variability}; \citealp{loreau1989coexistence}; \citealp{loreau1992time}; \citealp{klausmeier2010successional}; \citealp{li2016effects}; \citealp{letten2018species}; \citealp{schreiber2021positively}). Despite these counterexamples, the imprecision of the conventional interpretation is not widely recognized. Here we explain why the conventional interpretation is imprecise, even in models where a robust life-stage does engender a storage effect. Simply put, a robust life-stage does not prevent stochastic extinctions (i.e., extinction due to random chance, despite a positive invasion growth rate) nor does it improve the relative bet-hedging ability of rare species (i.e., the ability to increase fitness by decreasing the temporal variance of population growth).

Through examples, we show that $EC$ interaction effects arise readily from banal population dynamics. No special life-history adaptations are needed. In fact, we provide evidence for an interaction effect in a community of halophytic annual plants in Mediterranean grasslands of South Spain, leading to a storage effect that in no way depends on a persistent seed bank. We conclude that the storage effect is potentially everywhere, and that ecologists should employ expansive models that allow for the possibility of $EC$ interaction effects in every stage of species' life-cycles.

\section{The storage effect}
 \label{The storage effect}

Here we provide a brief description of the storage effect in order to ground our critique; experts may skip to the next section. For interested readers, a more comprehensive description of the storage effect is provided in Appendix \ref{A crash course in the storage effect}.

\subsection{The mathematical definition}
\label{The mathematical definition}

The mathematical definition of the storage effect is embedded within \textit{Modern Coexistence Theory} (\citealp{Chesson1994}; \citealp{chesson2000general}; \citealp{barabas2018chesson}), a framework for partitioning invasion growth rates into additive terms; these terms correspond to different explanations for coexistence, and are therefore called \textit{coexistence mechanisms}. All coexistence mechanisms are defined as a comparison between a rare species (the \textit{invader}) and species at their typical densities (the \textit{residents}). In a $S$-species community with residents $s$ and invader $i$, the storage effect is

\begin{equation} \label{SE math def}
    \Delta I_i = \zeta_i \;  \Cov{}{E_i}{C_i} - \sum \limits_{s \neq i}^S q_{is} \; \zeta_s \; \Cov{}{E_s}{C_s}.
\end{equation}

The parameter $E_j$ is called the \textit{environment}, the \textit{environmental parameter}, or the \textit{environment response}. It is typically a demographic parameter that depends on density-independent factors (e.g., germination probabilities and per capita seed production depends on precipitation), but can also represent the abiotic environment itself. The parameter $C_j$ is called the \textit{competition parameter}, but more generally represents the joint effects of density-dependent factors, which may include competitor densities, resources, predators, and mutualists. Note that $j$ is the index of an arbitrary species. Finally, the constants $q_{is}$, termed \textit{scaling factors}, scale residents' growth rates by a measure of relative sensitivity to competition (for all mathematical details, see Appendix \ref{A crash course in the storage effect}).

The coefficient $\zeta_j$ is the \textit{interaction effect of $E_j$ and $C_j$ on per capita growth rates}, defined as
\begin{equation} \label{zeta}
    \zeta_j = \frac{\partial^2 g_j(E_j^*,C_j^*)}{\partial E_j \partial C_j}, 
\end{equation}
where $g_j$ is the per capita growth function, which describes the average contribution of each individual to the growth of the population. The partial derivative is evaluated at the equilibrium parameter values $E_j^*$ and $C_j^*$, selected so that $g_j(E_j^*, C_j^*) = 0$. In continuous-time models, $g_j$ generates the per capita growth rate: $g_j(E_j(t), C_j(t)) = dn_j(t) / dt$. In discrete-time models, $g_j$ generates the effective per capita growth rate: the logged finite rate of increase, i.e., $g_j(E_j(t), C_j(t)) = \log(\lambda_j(t)) = \log(n_j(t+1) / n_j(t))$. 

% To give a brief example, the per capita growth rate function in the lottery model (\eqref{lottery}) can be written as  
% %
% \begin{equation}
%   g_j(E_j,C_j) = \log(s_j + \exp{E_j - C})
% \end{equation}
% %
% when the environmental and competition parameters are respectively defined as $E_j = \log(\eta_j)$ and $C = \log( \frac{\sum \limits_{k = 1}^{S} \eta_{k} n_{k} }{\sum \limits_{k = 1}^{S} (1-s_k) n_{k}})$. With this parameterization, the interaction effect is $\zeta_j = -s_j(1-s_j)$. Note that the interaction effect is zero when there is no adult survival.

\subsection{The ingredient-list definition}
\label{The ingredient list definition}

The storage effect depends on three ingredients:

\begin{enumerate}
    \item species-specific responses to the environment,
    
    \item a non-zero interaction effect with respect to fluctuations in the environment and competition (also known as \textit{nonadditivity} or an \textit{EC} interaction effect), and
    
    \item covariance between environment and competition (\textit{$EC$ covariance}).
    
\end{enumerate}

Ingredient \#1 --- species-specific responses to the environment --- simply establishes the presence of environmental niche differences, e.g., some species respond better to dry years vs. wet years. Ingredient \#2 --- an interaction effect --- is equivalent to the coefficient $\zeta_j$ (\eqref{zeta}). Ingredient \#3 --- the $EC$ covariance --- is generally satisfied when a favorable environment leads to high competition in the future, and when the environment does not change too quickly (\citealp{johnson2022towardsb}).

The storage effect generally has a positive effect on per capita growth rates (thus promoting coexistence) with a positive $EC$ covariance (Ingredient \#3) and negative interaction effect (Ingredient \#2), or with a negative $EC$ covariance and a positive interaction effect. Previous research has mainly focused on the former scenario, since a positive $EC$ covariance occurs readily (a good environment leads to high competition via intergenerational population buildup), though a negative $EC$ covariance (species are less sensitive to competition in favorable conditions) can arise in a negatively autocorrelated environment (\citealp{schreiber2021positively}). 

The interaction effect speaks to a \textit{synergy} between environment and competition: it is nor merely the case that a good environment leads to high competition and that high competition is bad for population growth; a negative interaction effect means that the simultaneous occurrence of a good environment and high competition is extra-bad.  Put another way, a negative (positive) interaction effect occurs when species are less (more) sensitive to competition in the face of a poor environment. For this reason, the negative interaction effect is sometimes referred to as \textit{buffering}; a positive interaction is referred to as \textit{amplifying}.

To demonstrate the association between the ingredients and coexistence, we consider a model with \textit{symmetric species} --- each species responds the environment in accordance with a symmetric covariance matrix with diagonal elements $\sigma^2$ and off-diagonal elements $\rho \sigma^2$, where $\rho$ is the between-species correlation in $E_j$. We assume that $\rho < 1$ (a statement  environmental niche differences) and that species are otherwise identical. In Appendix \ref{A crash course in the storage effect}, we show that the storage effect is
\begin{equation} \label{symmetry}
    \Delta I = - \zeta  (1 - \rho) \theta
\end{equation}
for every species. The three ingredients are captured  in the above formula: species-specific responses to the environment is $(1- \rho)$; the interaction effect is $\zeta$; and covariance is proportional to $\sigma^2 \theta$, where $\theta$ is a constant that converts the environmental responses of residents into competition. Mathematical expressions for the storage effect are generally more complicated in the non-symmetric case (e.g. Eq. 29 in \citealp{Chesson1994}).

\section{A critique of the conventional interpretation of the storage effect}
\label{A critique of the conventional interpretation of the storage effect}

Even though we can describe the storage effect using math, there remains a desire for a general ecological interpretation of the storage effect --- a concise, easy-to-understand explanation that links the phenomenon to well-known ecological constructs (e.g., stage-structure, dormancy, environmental niches). One such interpretation exists in the ecological milieu, as evidenced by 1) conversations with colleagues, 2) excerpts from the literature (Appendix \ref{The conventional interpretation: Excerpts from the literature}), and 3) the continued prominence of the lottery and annual plant models (\citealp{dean2020stochasticity}; \citealp{ellner2022toward}; \citealp{petry2018competition}; \citealp{zepeda2019fluctuation}; \citealp{bowler2022accounting}), wherein a robust life-stage is necessary for coexistence. This interpretation can be paraphrased as

\begin{quote}
\textbf{The conventional interpretation of the storage effect}: Species coexist by specializing on different parts of a fluctuating environment, so species must have a robust life stage in order to "wait it out" for a favorable time period. Thus, "storage" refers to a robust life stage can "wait it out".
\end{quote}

There are two separate problems with the conventional interpretation. First, in models where a robust life-stage is important for coexistence, the conventional interpretation implies that coexistence occurs because an invader is able to avoid stochastic extinction, or because the invader is employing a bet-hedging strategy; this is not true. Second, the conventional interpretation is not fully general. The storage effect arises readily in models without stage-structure, suggesting that a continued fixation on stage-structure will stymie the discovery of other routes to the storage effect.

Intuitively, "waiting it out" can help rare species avoid stochastic extinction. It is entirely reasonable to think that if species specialize on a fluctuating environment, they must have some way to slow the exponential loss of individuals over a sequence of bad years, and that this must be particularly important for rare species, which are inherently extinction-prone (\citealp{MacArthurRobertH1967Ttoi}; \citealp{lande1998demographic}). However important this phenomenon might be, it is not what the storage effect is measuring: most models used to demonstrate the storage effect feature infinite populations (an assumption made for mathematical/computational convenience), which obviates the possibility of stochastic extinction. An infinite population can lose an arbitrary number of individuals and still have infinite number of individuals left to lose. Unless per capita growth rates are $r = -\infty$ (which in most models is only possible as $n \rightarrow \infty$, a biological impossibility), extinction for infinite populations occurs asymptotically, i.e., after an infinite amount of time.

Alternatively, the idea of "waiting it out" smacks of \textit{bet-hedging}. In discrete-time and scalar-valued population models, an important quantity is $\lambda(t) = n(t+1)/n(t)$, known as the \textit{finite rate of increase} or \textit{ecological fitness}. Persistence is determine by the geometric mean of fitness (\citealp{Lewontin1969}; \citealp{Dempster1955}; \citealp{Stearns2000}; \citealp{Metz1992}), or equivalently, the sign of the effective average per capita growth rate, $\overline{\log(\lambda)}$. The average per capita growth rate can be approximated as $\overline{\log(\lambda)} \approx \overline{\lambda} - 1 - \frac{1}{2} \Var{}{\lambda}$, which reveals that species can benefit from adaptations that decrease the temporal variance of fitness, $\Var{}{\lambda}$, even if such adaptations incidentally decrease mean fitness $\overline{\lambda}$. Because these adaptations decrease the risk of catastrophic population decline, they are known as \textit{bet-hedging strategies}. 

Dormancy and iteroparous adults are widely-cited bet-hedging strategies (\citealp{cohen1966optimizing}; \citealp{rees1994delayed}; \citealp{venable2007bet}), so it would appear that the conventional interpretation is pointing at bet-hedging as the mediating mechanism of coexistence. In fact, the opposite is true. In Appendix \ref{bet_hedging}, we show that the storage effect does mediate bet-hedging, but tends to disproportionately reduce $\Var{}{\lambda_j}$ for resident species. This is because a negative $EC$ interaction effect reduces population growth (i.e., decreases variance) when the environment is favorable and competition is high, a context that is more likely to be experienced by resident species. A good environment for a resident species will lead to high competition, whereas a good environment for an invader will not. The rare-species disadvantage of bet-hedging is compensated by the fact that the storage effect (in total) disproportionately increases the invader's mean growth rate.

We have shown that stage-structure does not engender a storage effect through the suspected mechanisms (i.e., bet-hedging or avoiding stochastic extinction). Further, stage-structure is neither necessary nor sufficient for the storage effect. To see that a robust life stage is not sufficient, consider an arbitrary model with a robust life stage, but no environmental variation. A less trivial example is the modified lottery model wherein the survival probability fluctuates; here, the storage effect goes to zero as adults become more robust (Appendix \ref{Quantifying the storage effect in the lottery model}). A number of models have already demonstrated the possibility of the storage effect without stage structure (\citealp{abrams1984variability}; \citealp{loreau1989coexistence}; \citealp{loreau1992time}; \citealp{klausmeier2010successional}; \citealp{li2016effects}; \citealp{letten2018species}; \citealp{schreiber2021positively}). However, for purpose of illustration, we consider the following phytoplankton model,
\begin{equation} \label{phyto}
    \frac{d n_j(t)}{d t} = n_j(t) \left( b_j(t) R(t) - d_j \right),
\end{equation}
where $n_j(t)$ is the density of phytoplankton species $j$, $b_j(t)$ is the temporally-fluctuating uptake rate, $R(t)$ is nitrogen concentration, and $d_j$ is the death rate. Defining $E_j = \log(b_j)$ and $C_j = \log \left(-R\right)$, we find that the $EC$ interaction effect is $\zeta_j = -d_j$. Importantly, this interaction effect arises despite the lack of stage-structure. This phytoplankton model also demonstrates that an interaction effect, which either buffer or amplifies population growth, is not necessarily a result of life-history traits (e.g., dormancy, iteroparity) that have a clear adaptive purpose, but rather a by-product of banal features of population dynamics. The uptake rate is multiplied by the resource concentration, and the resulting multiplicative functional form of the per capita growth rate function gives rise to an interaction effect. 

Although the conventional interpretation is influential, experts in coexistence theory have long recognized that $EC$ interaction effects are not limited to systems with stage-structure. \citet{Chesson1994} writes "More generally, mechanisms leading a positive $\Delta I$ value [The storage effect] involve storage of the benefits of favorable periods in the population, whether this storage can be traced to a seed bank or something else. The term storage is a metaphor for the potential for periods of strong positive growth rate that cannot be canceled by negative growth at other times." This self-consciously abstract perspective lends itself to a more general interpretation of the storage effect, which we paraphrase as

\begin{quote}
\textbf{The conventional interpretation storage effect (v2)}: "Storage" can be more generally understood as \textit{buffering}, which is a negative interaction effect of environment and competition on per capita growth rates. This buffering helps out rare species because it prevents extreme losses when the environment is unfavorable and competition is high.
\end{quote}

Again, the existence of this interpretation is evidenced by excerpts from the literature (Appendix \ref{The conventional interpretation: Excerpts from the literature}). \textit{Buffering} is an apt way to describe a negative interaction effect, which truly does protect against the double whammy of a poor environment and high competition. However, the conventional interpretation v2 is imprecise because a species' storage effect tends to decrease a said species' buffering ability increases.

To explain further, we derive a mathematical expression for the storage effect in the two-species lottery model (Appendix \ref{Quantifying the storage effect in the lottery model}). The storage effect for the species 1 is proportional to $\left[ s_2 - \rho s_1 \right]$, where $s_1$ and $s_2$ are (respectively) the invader's and resident's adult survival probability. When species' responses to the environment are partially correlated (i.e., $0 < \rho < 1$), the storage effect decreases as the invader's adult survival probability increases. Since the invader's adult survival probability is measure of the invader's "storage" or "buffering", we observe that "storage" can decrease the storage effect. Note here that our critique assumes that the mathematical definition of the storage effect is the right way to define the storage effect; although alternative definitions are possible, they would not justify the conventional interpretation (v2) of the storage effect (Appendix \ref{An alternative definition of the storage effect}).

The seemingly paradoxical example of "storage" weakening the storage effect depends on species have positively correlated environmental responses. This is precisely what we expect to see in nature. It is well-known that plants have strong and positive growth responses to increases in temperature and precipitation (\citealp{rosenzweig1968net}; \citealp{lieth1973primary}; \citealp{sala1988primary}). The probability of germination --- which is often identified as the environmental response in models of annual plants --- can display a complex interdependency on temperature and precipitation, but nevertheless tends to increase as either abiotic variable increases (\citealp{baskin1998seeds}; \citealp{facelli2005differences}).

There is good empirical evidence that species have positively correlated environmental responses. We reviewed empirical studies that explicitly attempted to quantify or provide evidence for/against the storage effect (\citealp{caceres1997temporal}; \citealp{venable1993diversity}; \citealp{pake1995coexistence}; \citealp{pake1996seed}; \citealp{adler2006climate}; \citealp{sears2007new}; \citealp{descamps2005stable}; \citealp{angert2009functional}; \citealp{usinowicz2012coexistence}; \citealp{facelli2005differences}; \citealp{chesson2012storage}; \citealp{kelly2002coexistence}; \citealp{kelly2005new}; \citealp{usinowicz2017temporal}; \citealp{ignace2018role}; \citealp{hallett2019rainfall}; \citealp{armitage2019negative}; \citealp{armitage2020coexistence}; \citealp{zepeda2019fluctuation}; \citealp{zepeda2019fluctuation}; \citealp{towers2020requirements}; \citealp{jiang2007temperature}; \citealp{holt2014variation}; \citealp{ellner2016quantify}), including this paper's analysis of annual plant community (Section \ref{The Storage effect without a seed bank}).

In the 24 studies we were able to find, there were 16 distinct communities. Of these 16 communities, $8/16\; (50\%)$ showed evidence of positive correlations in species' environmental responses, $3/16\; (19\%)$ showed zero or near-zero correlation on average, $2/16\; (12\%)$ showed negative correlations, and $3/16\; (19\%)$ did not provide sufficient information to make a determination about the average sign of pairwise correlations. Two of the communities were only studied in the context of microcosm experiments. When we only consider natural communities for which sufficient information is available, $8/11\; (73\%)$ communities showed positive correlations and $2/11\; (18\%)$ show uncorrelated responses. Only $1$ community, a Mediterranean grassland (\citealp{hallett2019rainfall}), showed evidence of negative correlations. For more details on our analysis of the literature, see \textit{empirical\_E\_correlations.pdf} at \url{https://github.com/ejohnson6767/storage_effect_critique}.

To be clear, it \textit{is} true that in the lottery model (and the annual plant model), no species can have a positive storage effect if all species have zero adult (or seed) survival probability across time. However, \textit{this} is a community-level condition for coexistence, reflected properly in the ingredient list definition of the storage effect (see \ref{The ingredient list definition}). The conventional interpretation v2 may be thought of as conflating the community-level condition for coexistence (i.e., some species must have some "storage" for some species to coexist via the storage effect) with a species-level condition for persistence (i.e., one species must have "storage" in order for said species' storage effect to be positive). The conflation is analogous to falsely claiming that a species can persist by strongly competing with itself, since the competitive Lotka-Volterra model shows us that coexistence occurs when intraspecific competition is greater than interspecific competition.

\section{The Storage effect without a seed bank}
\label{The Storage effect without a seed bank}

The seminal models of coexistence theory --- the lottery and annual plant models --- have played a crucial role in the development of Modern Coexistence Theory. With just a little bit of biological realism, indeed with simple life-history traits, these models convincingly showed that environmental variation can (and likely does) promote coexistence in the real world. However, these models are very particular, whereas the storage effect is very general. We should not limit ourselves by only looking for the storage effect via the "robust life stage" mechanism.

The classic annual plant model (also called the seedbank model; \citealp{Chesson1994}) is written as
\begin{equation} \label{annual plants}
    X_j(t+1)= X_j(t) \left[ s_j (1 - G_j(t)) + \frac{G_j(t) Y_j}{1 + \sum_{k =1}^{S} c_k G_k(t) X_k(t)} \right], 
\end{equation}
where $X_j$ is the density of seeds of species $j$, $s_j$ is the probability that a seed survives the growing season if it does not germinate, $G_j(t)$ is the time-varying germination probability, $c_k$ are competition coefficients, and $Y_j$ is the maximum yield (seeds per germinant). Defining  $C_j(t) = C(t) = \log\left(1 + \sum_{k =1}^{S} c_k G_k(t) X_k(t)\right)$ and $E_j(t) = \log(G_j(t))$, the finite rate of increase can be written as 
\begin{equation} \label{annual plants2}
    \lambda_j(t) =  s_j (1 - G_j(t)) + \exp \left(E_j(t) - C(t)\right). 
\end{equation}
Using the mathematical recipe of \eqref{zeta}, we find that the interaction effect is $\zeta_j = -s_j$. The interaction effect is zero when there is no seedbank. To see this, imagine that $s_j = 0$, in which case the first additive term in \eqref{annual plants2} vanishes and the per capita growth rate becomes 
\begin{equation} \label{annual plants3}
   r_j(t) = \log(\lambda_j(t)) =  E_j(t) - C(t). 
\end{equation}
In this "no seed bank" scenario, it is easy to see that the parameters $E_j$ and $C_j$ have purely additive effects on the per capita growth rate --- there is no interaction effect. 

In general, the interaction effect will vanish for any growth rate function that takes the form $r = \log(\lambda) = \log(u(E) * v(C)) = \log(u(E)) + \log(v(C))$, where $u$ and $v$ are arbitrary but smooth functions. Put this way, the additivity of the annual plant model appears particular and unrealistic. In reality, the productivity in real annual plant communities is a complex function of events that occur over the length of the growing season, including size-dependent growth, the variability of precipitation, the dynamics of soil moisture, the timing of germination and flowering, etc. Additionally, there are several environmental parameters (e.g., soil moisture, nutrient content, herbivore and pollinator abundances). It is likely that all of this complexity harbours an $EC$ interaction effect. As \citealp{Chesson1994} writes, "There are so many ways in which nonadditivity can arise that it seems doubtful that any real populations could be additive, although approximate additivity could be common". 

%  From this more abstract perspective, the function for per capita seed production (i.e., the second additive term in \eqref{annual plants}) certainly seems particular. It is also unrealistic.

To illustrate the ubiquity of $EC$ interaction effects, we analyze a community of annual plants at Caracoles Ranch, in Doñana National Park, Spain. Data was collected for 19 plant species over 8 growing seasons (2015--2022), across a spatial extent of approximately $5000m^2$. The fruit production of individual plants (sample size = 11187)  was measured at peak fruiting time (i.e., when half of the flowers per individual have fruited), as well as the number and species identity of competitors within a 7.5 cm radius of the focal plants. Soil moisture was recorded at 2-week intervals at the spatial resolution of $1m^2$. The peak fruiting-time was highly variable across species and across years. For example, the Asteraceae \textit{Chamaemelum fuscatum} peaks on average in mid-April, whereas late phenology species with succulent leaves such as the succulent shrub Chenopodiaceae \textit{Salsola soda} peak in late September. Winter and spring precipitation highly influences the overall peak of fruiting across the community which varies from early May in very dry years (120 mm than average spring precipitation) to early July in wet years (90 mm than average spring precipitation).

% Defining competition as $\log(1+ \sum_k^S N_k)$, where $N_k$ is the number of germinants of species $k$ within a 7.5 cm radius of a focal individual. 

Because the of the high clay content in this particular area (77\%), soil moisture changes more across than within growing seasons, and such inter-annual variation appears to be determined by winter and spring precipitation. Therefore, we can define the environmental parameter $E$ as the temporal average of soil moisture throughout the duration of the growing season (January-May). Although soil moisture is ostensibly a density-dependent factor (plants remove water via evapotranspiration), soil moisture is surprisingly constant throughout the growing season (perhaps rainfall is intercepted and lost through evaporation) while still being highly predictive of seed production. Therefore, soil moisture can safely be treated as a density-dependent factor environmental parameter.

If the logarithm of fruit production, denoted $Z_j$, contains an $EC$ interaction effect, then the per capita growth rate will contain an $EC$ interaction effect, regardless of whether there is a seed bank. In addition to soil moisture, population growth is determined by the densities of nearby germinants, denoted $\boldsymbol{N} = (N_1, \ldots, N_S)^\intercal$. The full model, which takes the form of a multiple regression, contains nonlinear effects and an interaction effect:

% Soil moisture is highly predictive of productivity in arid environments (\citealp{??}). Indeed, in our annual plant system, soil moisture tends to correlate with fruit production in the absence of competition.

% \begin{equation}
%     r_j(t) = \log(\lambda_j(t)) = \log(G_j) + \log(\text{seeds per fruit}) + f_j(E_j, C_j)
% \end{equation}
%
\begin{equation} \label{fruit}
    Z_j = \beta_{0,j} + \beta_{1,j} E + \beta_{2,j} E^2 + \beta_{3,j} C_j + \beta_{4,j} C_j^2 + \beta_{5,j} E C_j + \epsilon_j, \quad \epsilon_j \sim \text{Normal}(0,\sigma_j)
\end{equation}

\begin{equation}
    C_j = \log \left(1 + \left(\sum_{k = 1}^S \left(\alpha_{jk} + \gamma_{jk} E \right) N_k \right)\right) 
\end{equation}
Here, the $\beta$'s are regression coefficients, $C$ is the competition parameter, $\sigma_j$ is the scale of residual variation. The "effective competition coefficient", defined as $\alpha_{jk} + \gamma_{jk} \times E$, is comprised of an intercept parameter and slope parameter. All parameters besides $\sigma_j$ have a hierarchical structure --- information is partially pooled across species in order to reduce estimation variance.

The functional form above is motivated by exploratory data analysis ($Z_j$ displayed an approximately linear relationship with $\log(1+\sum_k N_k)$ and $E$) and prior knowledge (soil moisture is highly predictive of productivity in semi-arid environments). The form and hierarchical structure of the model are also supported by model comparisons (Table \ref{loo}). Model-fitting was performed with the \textit{Stan} (\citealp{carpenter2017stan}) program in the \textit{R} software environment (\citealp{rcore2022}); more details are available in Appendix \ref{model fitting details}.

\begin{table}[ht]
\caption{Model comparisons. The acronym “$elpd$” stands for expected log predictive density. $\widehat{elpd}$ is an estimate of elpd, computed using Pareto Smoothed Importance Sampling, Leave-One-Out Cross Validation (PSIS LOO) as implemented by the \textit{loo} package (\citealp{vehtari2019loo}). $\Delta \widehat{elpd}$ is the difference with respect to the best-fit model. $SE \left(\Delta \widehat{elpd} \right)$ is the estimate of the standard error of $\Delta \widehat{elpd}$. The column "\# Parameters" does not include hyperparameters. The effective number of parameters is calculated as the difference between $\widehat{elpd}$ and the non-cross-validated log predictive density; this quantity is analogous to the bias correction term in AIC. \label{loo}}
\centering
\begin{tabular}{l l l l l}
  \hline \\
Description & $\Delta \widehat{elpd}$ & $SE\left(\Delta elpd \right)$ & \# Parameters & \# Effective parameters \\ 
  \midrule
\makecell{main, nonlinear, and interaction effects;\\$S^2$ competition coefficients;\\hierarchical} & 0.00 & 0.00 & 836.00 & 277.87 \\ 
  \makecell{main, nonlinear, and interaction effects;\\$S^2$ competition coefficients} & -20.74 & 11.77 & 836.00 & 300.38 \\ 
  \makecell{main and interaction effects;\\$S^2$ competition coefficients;\\hierarchical} & -111.89 & 17.94 & 798.00 & 249.48 \\ 
  \makecell{main and interaction effects;\\$S^2$ competition coefficients} & -155.74 & 19.98 & 798.00 & 291.65 \\ 
  \makecell{main effects;\\$S^2$ competition coefficients} & -325.45 & 28.72 & 779.00 & 205.93 \\ 
  \makecell{main effects;\\$S$ competition coefficients} & -898.15 & 45.41 & 95.00 & 84.04 \\ 
   \hline
\end{tabular}
\end{table}

% Finally, note here that $E$ is an abiotic variable, not a demographic parameter; this is slightly unconventional, but is not without precedent (e.g., \citealp{Ellner2019}).

\begin{figure} 
  \centering
      \includegraphics[scale = 0.6]{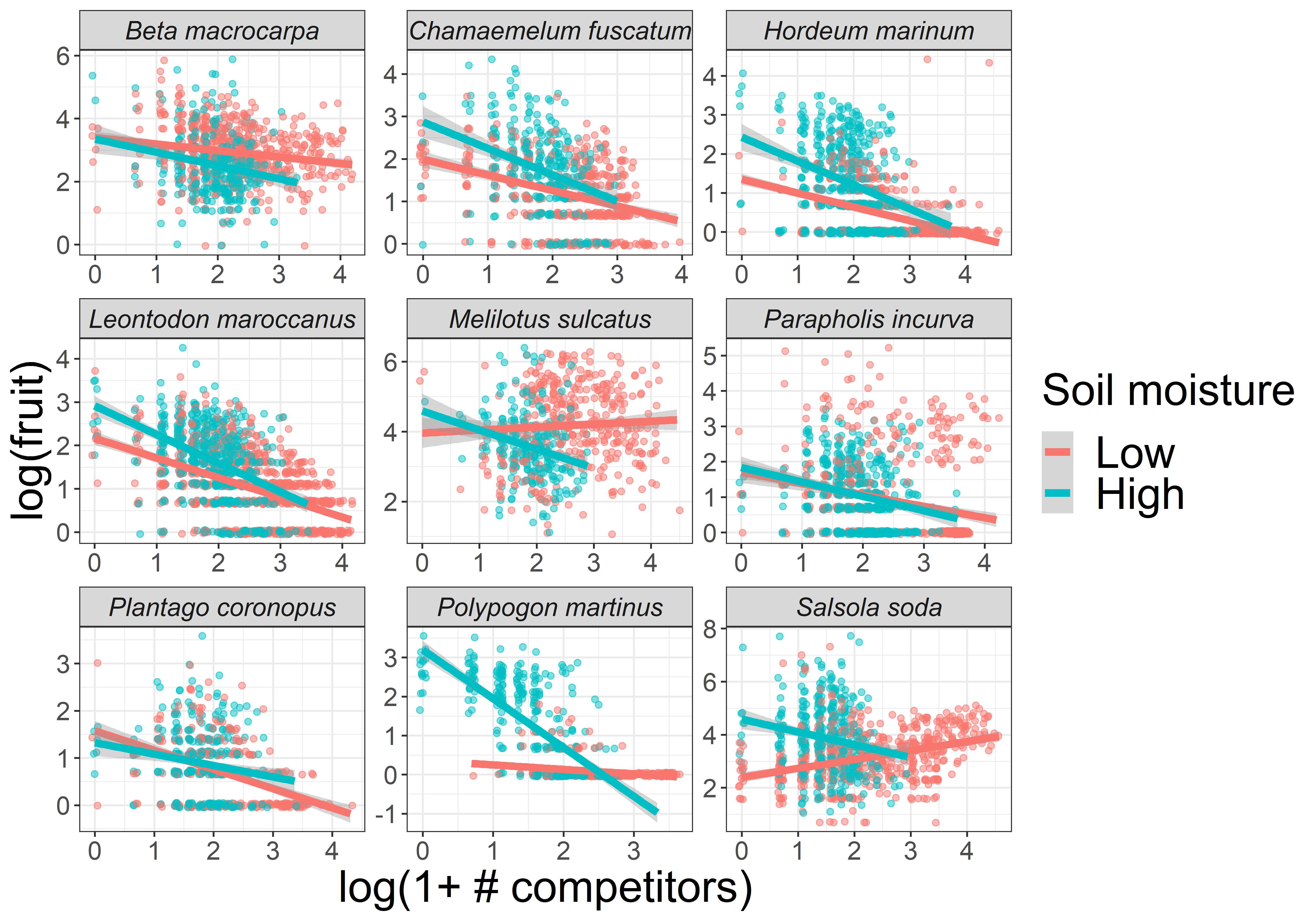}
  \caption{Graphical evidence of $EC$ interaction effects. The logarithm of fruit production, the proxy for the per capita growth rate, is plotted against $\log(1+\sum_k^S N_k)$, the proxy for competition. Soil moisture is considered low if it is below the mean. For all but one species, the slope is less negative in the low moisture regime. This pattern is indicative of a negative $EC$ interaction effect, also known as \textit{buffering} --- the  per capita growth rate is less sensitive to competition in the face of a poor environment. \label{interaction_effect}}
\end{figure}

\begin{figure} 
  \centering
      \includegraphics[scale = 0.6]{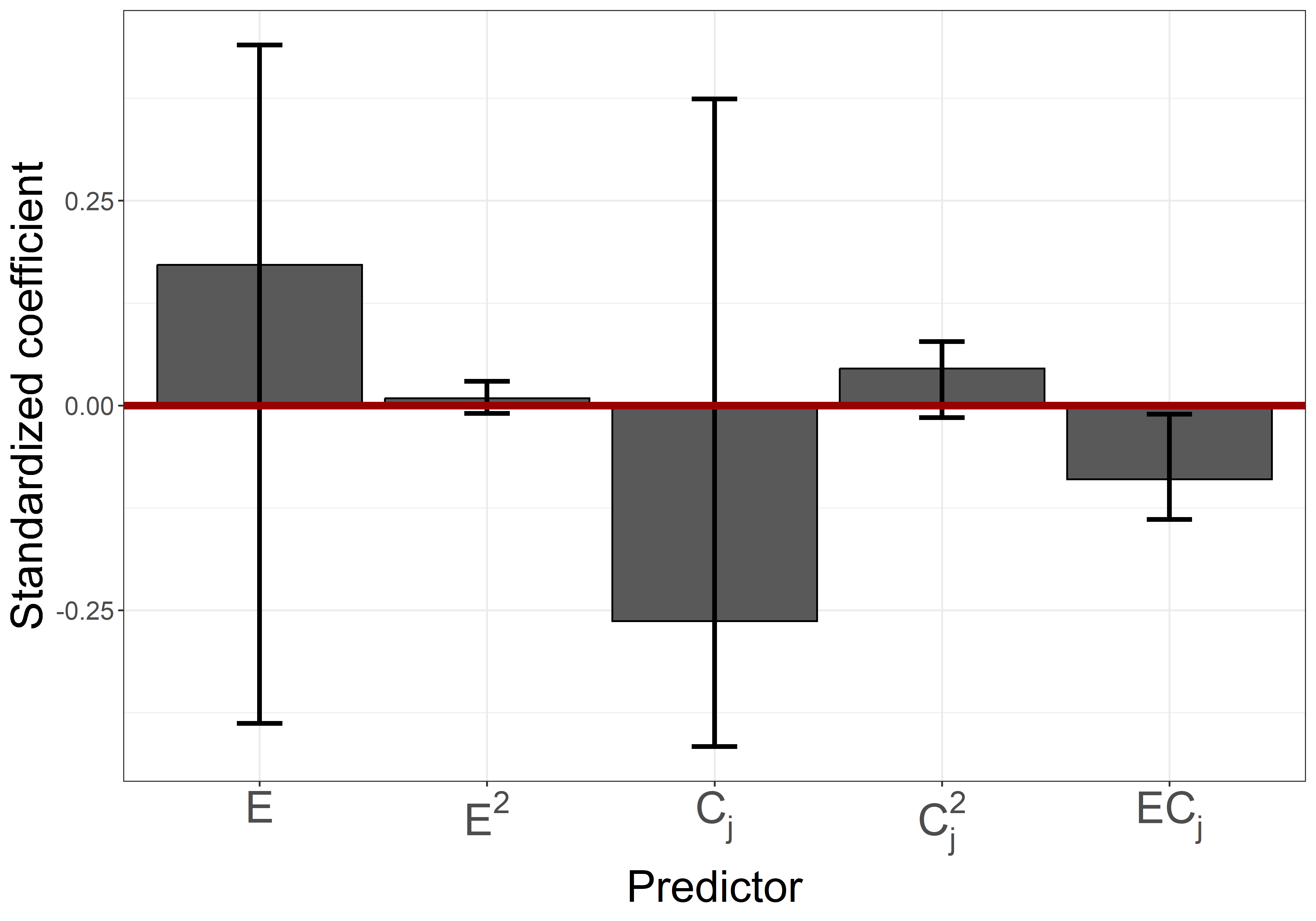}
  \caption{The standardized coefficient, a measure of predictor importance. The interaction effect is commensurate with the two main effects. The standardized coefficient is calculated as the posterior average of a model coefficient, divided by the standard deviation of the corresponding predictor. The bars and error bars represent the median and 50\% predictive intervals, respectively, across species.  \label{importance}}
\end{figure}

An $EC$ interaction effect is evident in Figure \ref{interaction_effect}: the slope of the $Z_j \sim \log(1+\sum_k N_j)$ regression is less negative in the low moisture regime. In other words, the effect of competition on the per capita growth rates becomes less severe in a poor environment, the hallmark of a negative interaction effect. Still, this graphical evidence assumes that all species exert the same competitive pressure. The multiple regression relaxes this assumption with the inclusion of pairwise competition coefficients, and confirms that the interaction effect, $\beta_{5,j}$, is commensurate in importance to the other regression coefficients (Fig. \ref{importance}). The effective per capita growth rate function is $\log(\lambda_j) = \log(\text{germination probability} \times \text{seeds per fruit}) + "\text{right-hand-side of $\eqref{fruit}$}"$, which implies that the $EC$ interaction effect is simply $\zeta_j = \beta_{5,j}$.

The model also provides evidence for the other ingredients: species-specific responses to the environment and $EC$ covariance (Fig. \ref{pairs}, \ref{response_hist}, \& \ref{cov}; Appendix \ref{model fitting details}). We do not quantify the storage effect, since this would require an analysis whose complexity exceeds the scope of this paper --- an analysis that parses spatial and temporal coexistence mechanisms, using a model that accounts for spatial heterogeneity, dispersal dynamics, and the dependence of germination and survival on moisture. However, the distributions of intra and interspecific competition coefficients are nearly identical (Fig. \ref{alpha}, Appendix \ref{model fitting details}), suggesting that fluctuation-dependent mechanisms are more important than classical coexistence mechanisms (i.e., resource/predator partitioning).

% However, we compute pseudo-coexistence mechanisms as "proof of concept". We assume that there is no seed survival, and that all species are demographically equivalent with the exception of species-specific responses to the environment, and differential intraspecific and interspecific competition coefficients. Under these idealized conditions, we see that that the storage effect promotes coexistence (Fig. \ref{??}), far more than classical coexistence mechanisms (i.e., resource/predator partitioning). We expect that this result extends to the real-world, given that the distribution of intraspecific competition coefficients is nearly identical to that of interspecific competition coefficients (Fig. \ref{alpha}, Appendix \ref{model fitting details}).

Where does the $EC$ interaction effect come from? We have no definitive answer, but we can offer some plausible explanations. In a year with high soil moisture, individual plants can grow large, and larger individuals produce stronger competition effects than smaller individuals (\citealp{rees2013competition}). In the face of high competition, at some point during the growing season, large plants stop being limited by soil moisture and start being limited by light or soil nutrients (\citealp{demalach2017light}). The presence of many large plants will undoubtedly intensify the negative effects of competition along some dimension, resulting in a negative $EC$ interaction effect. To back up our verbal argument, we present a logistic model for the within-generation dynamics of size. Plant size at day $s$ of the growing season is $x(s)$. The number of germinants is $N$, and soil moisture is $E$. Integrating the logistic model, 
\begin{equation}
\frac{d x(s)}{d s} = x(s) E (1- a N x(s)),
\end{equation}
from $0$ to $h$, the duration of the growing season, gives us plant size. Now, if we define competition as $C_j = a N x(h)$, and claim that seed production is proportional to plant size, then the finite rate of increase can be written as $\lambda = C + (1-C)\exp(E h)$. Applying the definition of the interaction effect (\eqref{interaction_effect}), we obtain a negative interaction effect, $\zeta_j = - h \exp(h E^*)$.

\section{Discussion}
\label{Discussion: Rethinking storage}

This paper has two main messages. First, when a robust life stage does engender a storage effect, it does not do so via the suspected mechanisms --- by preventing stochastic extinction or a rare-species advantage in bet-hedging. Rather, a robust life stage can engender a storage effect by enabling an interaction effect between environment and competition (item \#2 in the ingredient-list definition of the storage effect; Section \ref{The ingredient list definition}). Second, such $EC$ interaction effects may arise from a variety of processes, so ecologists should search for the storage effect in a phenomenological way, and later provide a mechanistic understanding, if possible. 

% The storage effect is an important concept in community ecology, but it remains widely misunderstood. We have shown that simple interpretations fail, leaving us in the dark; 

The question remains, how should the storage effect be understood? Our suggestion is that understanding should be built around the ingredient list definition of the storage effect (Section \ref{A crash course in the storage effect}), which states that the storage effect tends to support many species when there are 1) species-specific responses to the environment, 2) interaction effects between environment and competition; and 3) covariances between environment and competition. The presence of the three ingredients can be checked with exploratory plots (e.g., Fig. \ref{interaction_effect}, \ref{pairs}). A holistic understanding of the storage effect can be achieved by relating the ingredient-list definition, the mathematical definition, and concrete examples; indeed, this is the project attempted by Appendix \ref{A crash course in the storage effect}. 

Two of the three ingredients of the storage effects are straightforward to interpret. Ingredient \#1 (species-specific responses to the environment) is simply a type of niche partitioning. Ingredient \#3 ($EC$ covariance) is generically fulfilled when a good environment leads to population growth, and subsequently, competition via overcrowding (\citealp{johnson2022towardsb}). Ingredient \#2, an $EC$ interaction effect, which is the subject of this paper, requires a more abstract interpretation. It is the \textit{synergistic (or antagonistic) effect} of environment $E$ and competition $C$; it is the degree to which a favorable $E$ exacerbates (or alleviates) the effects of $C$. A negative interaction effect can be interpreted as protection against the double-whammy of a poor $E$ and high $C$, whereas a positive interaction effect can be interpreted as acute susceptibility to a poor $E$ and high $C$

% An $EC$ interaction effect can also be interpreted as producing an "effective regulating factor". Coexistence require given a functional interpretation: coexistence requires specialization on regulating factors, so the $EC$ interaction effect combines the specialization inherent in species' environmental responses, with the density-dependence inherent in the competition parameter, thus functionally creating an "effective regulating factor". 

Although the consequences of the $EC$ interaction effect for the population dynamics of competing species can be understood, a unique ecological interpretation does not exist --- interaction effects are ubiquitous and therefore cannot be tied to any particular mechanism. This acknowledgement of complexity should resonate with both quantitative ecologists (the regression coefficients are never zero) and environmentalists ("everything is connected to everything else", \citealp{commoner2015closing}). Of course, in particular models, we can use the abstract interpretations (from the previous paragraph) to hypothesize about particular mechanisms of an $EC$ interaction effect, but our imaginations need not limit our inferences. When looking for the storage effect, ecologists should look for interaction effects wherever possible.

% , and the same occur for explaining the nature of self-limiting processes, interspecific interactions, or higher-order interactions in multispecies settings. 

When exploring ecological processes using a phenomenological perspective, one might worry that the inclusion of interaction effects will increase the risk of overfitting (\citealp{hastie2009elements}, Ch. 7), but overfitting can be monitored with Cross Validation and abated with regularization (i.e., methods for reducing estimation variance in many-parameter models). For example, the best-fit model of our annual plant system has 836 parameters, but prior distributions and hierarchical model structure reduces the number of effective parameters to 278 (a 66\% reduction in complexity). Additional variance-reduction techniques could be employed, including model-averaging (\cite{dormann2018model}), sparsity-inducing priors (\citealp{carvalho2009handling}) and sub-models for  constraining competition coefficients (\citealp{weiss2022disentangling}).

% It is possible that much of the confusion around the storage effect can be explained by a psychological tendency towards explaining coexistence in terms of how a single species is able to recover from rarity. We speculate that this tendency exists for the same reason that ecologists use invasion analyses: it breaks up an abstract property --- coexistence --- into a set of simpler, concrete scenarios (i.e., can a species invade in the context determined by the limiting dynamics of resident species?). 

\begin{figure} 
  \centering
      \includegraphics[scale = 0.45]{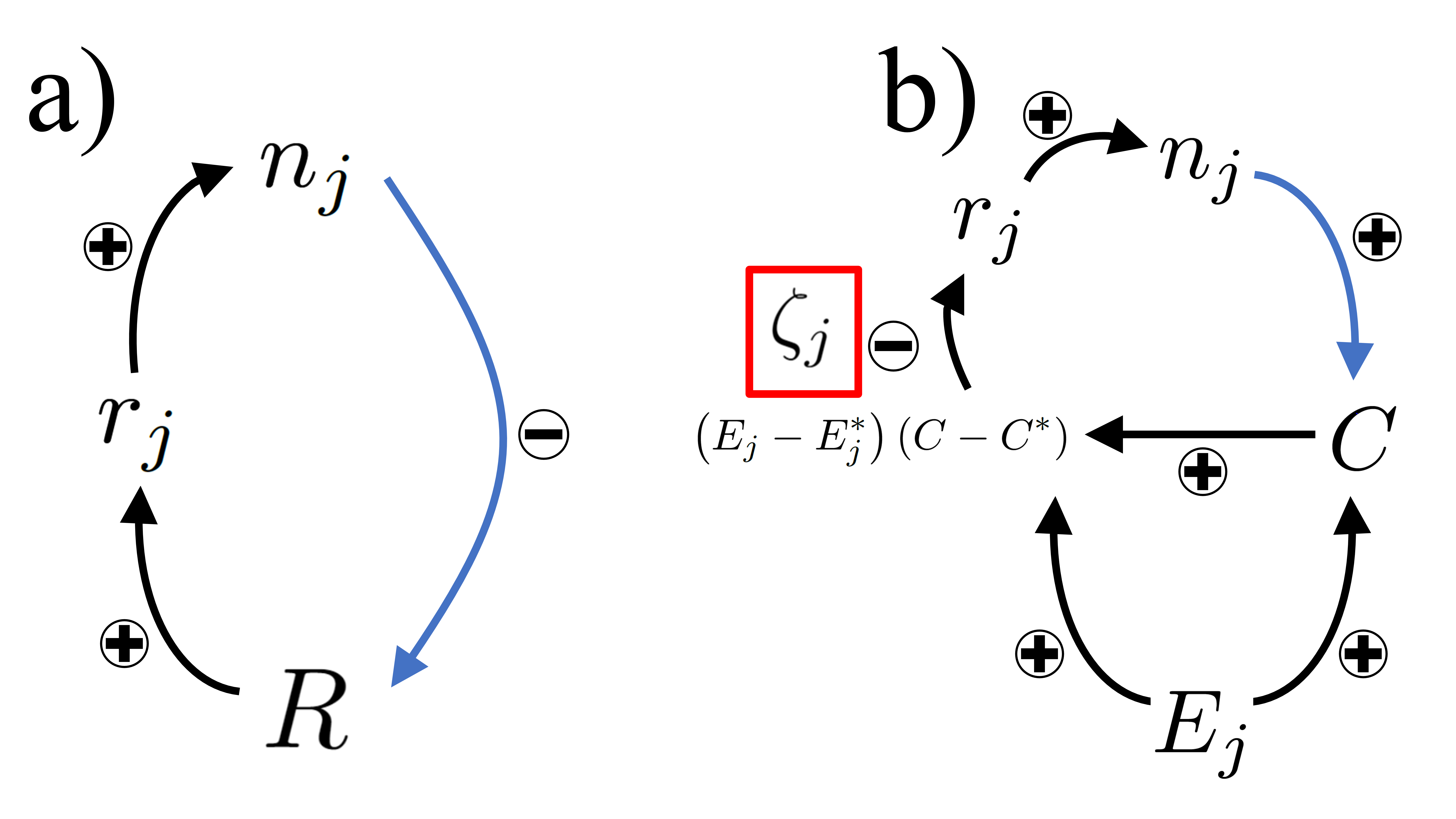}
  \caption{Density-dependent feedback loops for species that coexist due to resource partitioning (Panel a) and the storage effect (Panel b). The diagrams depict the self-limitation of a single species (subscript $j$), a subset of a much larger community-level causal diagram. The blue arrows vanish when a species is perturbed to the invader state: zero density. \textbf{Panel a)}: population density $n_j$ has negative effects on resource concentration $R$, and resource concentration has a positive effect on the per capita growth rate $r_j$.  \textbf{Panel b)}: population density has a positive effect on the competition parameter. Thus, the population indirectly has a positive effect on the product of fluctuations $(E_j - E_j^*)(C_j - C_j^*)$ (this quantity becomes the $EC$ covariance when averaged across time). The product of fluctuations has a negative effect on the per capita growth rate if the interaction effect $\zeta_j$ (red box) is negative. \label{causal_diag}}
\end{figure}

When formulating heuristic explanations for coexistence, one naturally considers a scenario in which a single species is rare, and then asks what allows this species to recover from rarity. In the case of coexistence via classical mechanisms, the heuristic explanation is, "If a species were ever to become rare, the resource that it specializes on will become more abundant, thus increasing per capita growth rates" (Fig. \ref{causal_diag}a). Here, the density-dependent feedback loop contains intuitive state variables (i.e., species densities and resource concentrations, specifically their mean levels) which interact in an obvious way (i.e., more resources per capita $\rightarrow$ higher per capita growth rates). The analogous heuristic explanation for coexistence via the storage effect is "If a species were ever to become rare, then the environmental states it specializes on will not lead to as much competition (as they did previously), the covariance between environment and competition will decrease, thus increasing per capita growth rates" (Fig. \ref{causal_diag}b). Here, the density-dependent feedback loop contains an abstract state variable (i.e., the product of fluctuations, $(E_j - E_j^*)(C_j - C_j^*)$) that interacts with a species' density in a non-obvious way (i.e., a negative interaction effect $\zeta_j$, deduced via a Taylor series). This explanation is unintuitive and requires an understanding of the math behind the storage effect. It is simply easier to say that rare species "store good years of recruitment" or "are buffered against unfavorable environmental conditions". But this is incorrect: the infinitesimal density of rare species means that any addition to density (i.e., "storage" in the conventional sense) has no effect on population dynamics, and "buffering" (as is it defined mathematically) tends to hurt rare species.

% The failure of the species-centric interpretations implies that
The storage effect is best understood in a community context. The ingredient-list definition of the storage effect gives the community-level "conditions", in the sense that the ingredients \textit{tend} to lead to a positive (or negative) storage effect for many species. Even though the species-specific $\Delta I_i$ (\eqref{SE math def}) is called "the storage effect" by convention, a number of papers have identified the community-average measure as the more relevant quantity (\citealp{chesson2003quantifying}; \citealp{Chesson2008quantifying}; \citealp{Yuan2015}).

The generality of the storage effect is both a strength and weakness. It is a weakness because a complete interpretation cannot rely on well-known concepts like bet-hedging, or a small set of life-history traits like dormancy or robust adults. Instead, the storage effect requires a phenomenological explanation through abstractions like "interaction effects" and "$EC$ covariance". Nevertheless, the generality of the storage effect is a strength because it gives us a way to analyze, talk about, and quantify a phenomenon that occurs in disparate systems.

\newpage{}

\section*{Acknowledgements}
We would like to thank Karen Abbott for helpful suggestions. OG acknowledges financial support provided by the Spanish Ministry of Science and Innovation (PID2021-127607OB-100) and through Ramón y Cajal programm (RYC-2017-23666).

\section*{Author contribution statement}
E.J. conceived the project and wrote the first draft; O.G. and collaborators collected the annual plant data; E.J. analyzed the data; A.H. \& O.G. contributed substantial revisions.

\section*{Data availability statement}
All pertinent files and code will be available at \url{https://github.com/ejohnson6767/storage_effect_critique}.

\section*{Conflict of interest statement}
The authors declare no conflicts of interest.

\newpage{}

\begin{appendices}

\section{Evidence of the conventional interpretation: Excerpts from the literature}
\label{The conventional interpretation: Excerpts from the literature}

Here we attempt to show that the conventional interpretation exists. We are not making any claims about what the quoted authors do or do not know about the storage effect --- it is often pedagogically useful to present definitions that are evocative but not 100\% precise. Similarly, there is no sense in giving a fully general account of the storage effect when discussing how the phenomena emerges from a particular model. Our only purpose in providing these quotations is to show that a reasonable reader could distill the conventional interpretation (or something like it) from the literature.  

\subsection*{Evidence of the conventional interpretation}

\begin{itemize}

\item "\ldots adults must be able to survive over periods of poor recruitment, such that the population declines only slowly during these periods. Under these conditions, a species tends to recover from low densities, and competitive exclusion is opposed. .... We refer to this phenomenon as the storage effect because strong recruitments are essentially stored in the adult population, and are capable of contributing to reproduction when favorable conditions return (\citealp{chesson1985coexistence}).

\item "\ldots the storage-effect coexistence mechanism relies on such buffering effects of persistent stages, because these prevent catastrophic population decline when poor recruitment occurs." (\citealp{chesson2003quantifying})

\item "Persistence of adults limits the damage from unfavourable conditions, but does not prevent strong growth at other times. \ldots Similarly, the dormant seeds of annual plants are relatively insensitive to environmental factors and competition in comparison with the actively growing plants." (\citealp{Chesson2004})

\item "Seed banks or long-lived adults ‘‘store’’ the effects of favorable years, which buffer the effects of bad years when population sizes may decline." (\citealp{sears2007new})

\item "First, organisms must have some mechanism for persisting during unfavourable periods, such as a seedbank, quiescence or diapause. This condition, which gives the storage effect its name, buffers negative population growth; without it, populations would go extinct after a brief unfavourable period and environmental variation could never promote coexistence." (\citealp{adler2014testing})

\end{itemize}

\subsection*{Evidence of the conventional interpretation (v2)}

\begin{itemize}

    \item "\ldots there is some way to “store” the effects of good times, to get organisms through bad ones." (\citealp{barabas2018chesson})
    
    \item "More generally, mechanisms leading a positive $\Delta I$ value involve storage of the benefits of favorable periods in the population, whether this storage can be traced to a seed bank of something else. The term storage is a metaphor for the potential for periods of strong positive growth rate that cannot be canceled by negative growth at other times." (\citealp{Chesson1994})
    
    \item "Storage effects happen when the invader experiences low competition in favorable environments and has the ability to store that double benefit." (\citealp{snyder2012storage})

    \item "However, these gains by the rare come to nothing if they are wiped out in bad years. The storage effect can therefore maintain coexistence only if species are buffered against sudden rapid declines. One natural way for this to occur is if generations overlap and established individuals are immune to the causes of temporal variation (e.g., viability selection on offspring, no selection on adults)." (\citealp{messer2016can})

\end{itemize}

\section{An extended description of the storage effect}
 \label{A crash course in the storage effect}

\subsection{An introductory example: The lottery model}

The storage effect is well-demonstrated with a toy model of coral reef fish dynamics. The \textit{lottery hypothesis} (\citealp{Sale1977}) states that the local diversity of coral reef fishes is generated by the random allocation of space: when an adult fish dies, the various fish species enter a lottery for the open territory with a number of tickets equal to the number of larvae that each fish species produces. The lottery hypothesis was motivated by the fact that coral reef fishes do not appear to finely partition food types, but do appear to be limited by space. Space limitation is evidenced by the observed territoriality of adults (\citealp{warner1980population}), the production of larvae in massive numbers, and the weak correlation between adult population size and the subsequent number of recruits (\citealp{cushing1971dependence}, \citealp{szuwalski2015examining}). 

While Sale's lottery hypothesis does a fine job at explaining local biodiversity, it cannot explain the maintenance of biodiversity --- coexistence. \citet{chesson1981environmentalST} were able to attain coexistence with the addition of a single feature: temporal variability in per capita larval production. The resulting model is now known as the \textit{lottery model}, and the more general process permitting coexistence is known as \textit{the storage effect}. The exposition here follows the lottery model of \citet{Chesson1994}, as opposed to original lottery model (\citealp{chesson1981environmentalST}), which is more complex due to stochasticity in both adult mortality and larval production.

Imagine a guild of fish species inhabiting discrete territories on a coral reef. Several events occur in each time-step of the lottery model, here presented in chronological order

\begin{enumerate}
    \item The fish spawn. Per capita larval-production, (i.e., per capita fecundity) fluctuates from time-step to time-step, putatively due to dependency on environmental factors that also fluctuate. Like the larvae of many marine fish, our hypothetical larvae disperse offshore (ostensibly to avoid predation) and return some time later, though still within the time-step.
    
    \item Adult fish die with some density-independent probability, leaving behind an empty territory. The death probability may vary across species, but unlike fecundity, does not vary across time.
    
    \item The larvae return to the reef and inherit the empty territories with a recruitment probability for any given larva being equal to the number of empty sites, divided by the total number of larvae. This uniform probability of per larva recruitment is the \textit{lottery} in the lottery model. The unrecruited larvae die before the next time-step begins. 

\end{enumerate}

The above dynamics are expressed in the difference equations,

\begin{equation} \label{lottery}
\begin{aligned}
   n_j(t+1) =  n_j(t) \left[\overbrace{s_j}^{\text{survival prob.}} + \; \overbrace{\eta_{j}(t)}^{{\scriptstyle \text{per capita fecundity}}} \; \left[\rule{0cm}{1.25cm}\right. \frac{ \overbrace{ \sum \limits_{j \neq i}^{S} (1-s_j) n_{j}(t)}^{\text{open territories}}}{ \underbrace{\sum \limits_{j \neq i}^{S} \eta_{j}(t) n_{j}(t)}_{ {\scriptstyle \text{total larvae}}}} \left. \rule{0cm}{1.25cm}\right] \right], \quad j = (1, 2, \ldots, S), 
   \end{aligned}
\end{equation}

where $n_j(t)$ is the density of species $j$ at time $t$, $s_j$ is the adult survival probability, and $\eta_j(t)$ is the time-varying per capita larval production.

\begin{figure} 
  \centering
      \includegraphics[scale = 1.2]{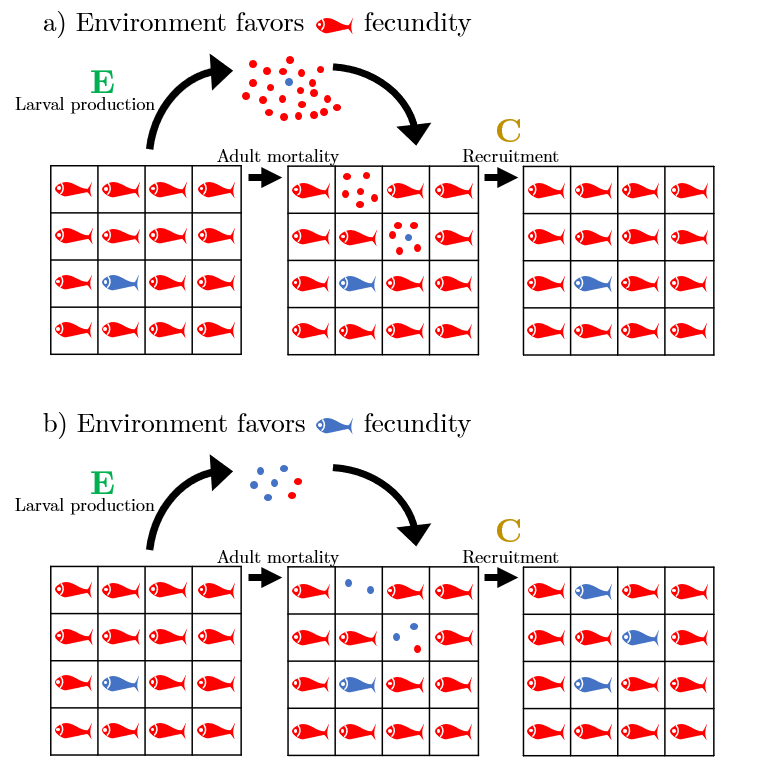}
  \caption{Diagrammatic representation of the lottery model\label{lottery model}. Figure originally appears in \citeyear{johnson2022towardsb}}
\end{figure}

In the lottery model, empty space is the only limiting resource, so the competitive exclusion principle (which states that no more than $N$ species can coexist on $N$ regulating factors; (\citealp{volterra1926variationsST}, \citealp{Lotka1932TheSupply}, \citealp{levin1970community}) is transcended if even two fish species are able to coexist. Intuitively, if a species becomes rare for whatever reason (e.g., competition, catastrophes), then it must have a positive per capita growth rate if it is to recover from rarity. This is a slight simplification (see \citealp{barabas2018chesson}, p. 293), but for our current purposes, we say that coexistence is related to a \textit{rare-species advantage}, operationalized by the \textit{invasion growth rate}: the long-term average per capita growth rate of a species that has been perturbed to low density. The practice of determining coexistence based on invasion growth rates is called an \textit{invasion analysis} (\citealp{turelli1978reexamination}; \citealp{Grainger2019TheResearch}). 

Consider a two-species lottery model with a red species and a blue species  (Fig. \ref{lottery model}). The red species produces many larvae during hot years and few larvae during cold years. The blue species responds to temperature oppositely: it produces few larvae during hot years and many larvae during cold years. We now ask the question pertinent to coexistence: if the blue fish species becomes rare, will it be able to recover? 

When the blue fish species experiences a good (i.e., cold) year, there are few larvae produced in total: the environment is unfavorable to the red fish, and although there are many blue larvae produced per capita, the blue fish are rare. Each blue fish larva thus experiences relatively little competition, which we may measure as larvae per empty site. In this scenario, we see that a rare species is able to capitalize on a good environment (Fig. \ref{lottery model}a).

To uncover a potential rare-species advantage, we must now examine an analogous scenario from the perspective of the common species. When the red fish species experiences a good (i.e., warm) year, many red larvae are produced in total, since there are many red fish and the environment favors the red fish. However, there is now an excess of red larvae, which significantly decreases the probability of any one larva winning a territory. The consequence of this high competition is a small or zero-valued per capita growth rate; the last panel of Fig. \ref{lottery model}b shows no net change). Unlike the rare blue fish, the common red fish is unable to capitalize on a good environment. 

We will now frame this concrete scenario (i.e., the red fish experiencing a hot year) in slightly more general terms: For a common species, a good response to the environment (e.g., high per capita larval fecundity) causes high competition (e.g., many larvae per open site), which ultimately undermines the good response to the environment. For a rare species, a good response to the environment does not lead to as much competition. It is this asymmetry between rare and common species that drives the storage effect.

\subsection{A simple interpretation of the storage effect}
\label{An okay interpretation of the storage effect}

\begin{siderules}
Good environments lead to high competition for common species, but less so for rare species. Since high competition undermines the positive effects of a good environment (via a negative interaction effect of environment and competition on per capita growth rates), rare species are better able to take advantage of a good environment than common species.
\end{siderules}

This interpretation is correct but leaves out some details. How can the asymmetry between rare species and common species be represented mathematically? For that matter, how can the negative interaction effect be represented mathematically? Is the storage effect a species-level or community-level characteristic? These questions are best answered with the mathematical definition of the storage effect and its textual analogue: an ingredient list of conditions for the storage effect.

\subsection{Modern Coexistence Theory}
\label{Modern Coexistence Theory}

The mathematical definition of the storage effect is embedded within \textit{Modern Coexistence Theory} (\citealp{Chesson1994}; \citealp{chesson2000general}; \citealp{barabas2018chesson}), a framework for partitioning invasion growth rates into additive terms; these terms correspond to different explanations for coexistence, and  are therefore called \textit{coexistence mechanisms}. The storage effect is one of several coexistence mechanisms. In order to arrive at the mathematical definition of the storage effect, we provide a step-wise summary of the derivation of coexistence mechanisms:

\begin{enumerate}
    \item \textbf{Choose one species to be the rare species}. This species is called the \textit{invader} and the remaining species are called \textit{residents}. Set the invader's density to zero, and let residents attain their limiting dynamics, i.e., let them equilibrate to their typical densities.
    
    The invader is denoted by the subscript $i$, the residents are denoted by the subscript $s$, and a generic species is denoted by the subscript $j$.
    
    \item \textbf{Write the per capita growth rates in terms of the environmental and competition}. Let the per capita growth rate of species $j$ be some function $g_j$ of the environmental parameter $E_j(t)$ and competition $C_j(t)$, i.e., $dn_j(t)/(n_j(t) dt) = r_j(t) = g_j(E_j(t),C_j(t))$. In discrete-time models like the lottery model, the effective per capita growth rate is the logged \textit{finite rate of increase}, i.e., $r_j(t) = \log(\lambda_j(t))$, where $\lambda_j(t) = n_j(t+1)/n_j(t)$. Extensions to structured populations can be found in \citet{Ellner2019}. For notational simplicity, we drop the explicit dependence on time $t$.

    The parameter $E_j$ is called the environmental parameter, the environmentally-dependent parameter, or simply the environment. It is typically a demographic parameter, belonging to species $j$, that depends on the abiotic environment (e.g., germination probability depends on temperature). More generally, $E_j$ may represent the effects of density-independent factors. The parameter $C_j$ is called the competition parameter, but it may represent the effects of density-dependent factors. Of particular interest is the case where $C_j$ is a function of shared predators, potentially leading to the storage effect due to predation (\citealp{kuang2010interacting}; \citealp{chesson2010storage}; \citealp{stump2017optimally}). 

    The invasion growth rate is the long-term average per capita growth rate of the invader. More generally, the invasion growth rate is the dominant Lyapunov exponent of the dynamical system representing population dynamics (\citealp{Metz1992}; \citealp{dennis2003can}).

    \item \textbf{Expand growth rates with respect to $E_j$ and $C_j$}. First, select equilibrium values of the environment and competition, $E_j^*$ and $C_j^*$, such that $g_j(E_j^*, C_j^*) = 0$. Next, perform a second-order Taylor series expansion of $g_j(E_j, C_j)$ about $E_j^*$ and $C_j^*$. 

    The result is

\begin{equation} \label{taylor_decomp}
\begin{aligned}
 {r_{j}(E_j,C_j)} \approx \; & \alpha_j^{(1)} (E_j - E_{j}^{*}) + \beta_j^{(1)} (C_j - C_{j}^{*}) \\ & + 
\frac{1}{2} \alpha_j^{(2)} (E_j - E_{j}^{*})^{2} + \frac{1}{2} \beta_j^{(2)} (C_j - C_{j}^{*})^{2} + 
\zeta_j  (E_j - E_{j}^{*})   (C_j - C_{j}^{*}),
\end{aligned}
\end{equation}

where the coefficients of the Taylor series, 

\begin{equation}  \label{taylor_coef}
\begin{aligned}
 \alpha_j^{(1)} = \pdv{g_j\scriptstyle{(E_j^*, C_j^*)}}{E_j},  \quad
 \beta_j^{(1)} = \pdv{g_j\scriptstyle{(E_j^*, C_j^*)}}{C_j},  \quad
 \alpha_j^{(2)} = \pdv[2]{g_j\scriptstyle{(E_j^*, C_j^*)}}{E_j},  \quad
 \beta_j^{(2)} = \pdv[2]{g_j\scriptstyle{(E_j^*, C_j^*)}}{C_j,}  \quad
 \zeta_j = \pdv{g\scriptstyle{(E_j^*, C_j^*)}}{E_j}{C_j},  \quad
\end{aligned}
\end{equation}

are all evaluated at $E_j = E_j^*$ and $C_j = C_j^*$.

\item \textbf{Time-averaging}. Invasibility is determined by what happens in the long-run, so our next step is to take the temporal average of \eqref{taylor_decomp}. Temporal averages are denoted with "bars"; e.g., the average per capita growth rate of species $j$ is $\overline{r}_j$.

\begin{equation} \label{taylor_decomp_avg1}
\begin{aligned}
\overline{r}_{j} \approx \; & \alpha_j^{(1)} (\overline{E_j} - E_{j}^{*}) + \beta_j^{(1)} (\overline{C_j} - C_{j}^{*}) \\ & + 
\frac{1}{2} \alpha_j^{(2)} \Var{}{E_j} + \frac{1}{2} \beta_j^{(2)} \Var{}{C_j} + 
\zeta_j  \Cov{}{E_j}{C_j}.
\end{aligned}
\end{equation}

The above expression above rests on several assumptions about the magnitude of environmental fluctuations and the relationship between environment, population density, and competition (details can be found in \citet{Chesson1994} and \citet{chesson2000general}). Most crucially, we assume that environmental fluctuations, $\abs{E_j - E_j^*}$, are very small, and that average environmental fluctuations  $\abs{\overline{E_j} - E_j^*}$, are even smaller. These \textit{small-noise assumptions} ensure that the expression \eqref{taylor_decomp_avg1} is a good approximation of the true invasion growth rate, thus justifying the truncation of the Taylor series at second order. The small-noise assumptions also justify the replacement of second-order polynomial terms with central moments, e.g., $\overline{(E_j - E_j^*)^2}$ is replaced by $\Var{}{E_j}$: the equilibrium value $E_j^*$ is assumed to be very close the temporal average $\overline{E_j}$, such that little growth rate is lost by replacing the former with the latter.

\item \textbf{Invader--resident comparisons}

The long-term average growth rate of each resident must be zero (otherwise residents would go extinct or explode to infinity), so the value of the invasion growth rate is unaltered if we subtract a linear combination of the residents' long-term average growth rates. 

\begin{equation} \label{inv res comparison}
    \overline{r}_i = \overline{r}_i - \sum \limits_{s \neq i}^S q_{is}  \overline{r}_s
\end{equation}

The $q_{is}$ are called \textit{scaling factors} (\citealp{barabas2018chesson}) or comparison quotients (\citealp{Chesson2019}). \posscite{Chesson1994} original definition of the $q_{is}$ utilized the so-called \textit{standard parameters} (to be discussed shortly), but is essentially equivalent to $q_{is} = \frac{\beta_i^{(1)}}{\beta_s^{(1)}} \frac{\partial C_i}{\partial C_s}$. Ellner et al. (\citeyear{ellner2016quantify}, \citeyear{Ellner2019}) have suggested scaling resident growth rates to create a simple average over resident species, i.e., in \eqref{inv res comparison}, replace the $q_{is}$ with $1/(S-1)$. We (\citealp{johnson2022methodsb}) have argued in favor of replacing the scaling factors with quotients of species' generation times. For the sake of convention, however, we use Chesson's original scaling factors.

Though the average growth rate of each resident is zero, the components of the average growth rate (i.e., the additive terms in \eqref{taylor_decomp_avg1}) are not necessarily zero. Therefore, we can draw meaningful comparisons between the invader and the residents by substituting the right-hand side of the Taylor series expansion (\eqref{taylor_decomp_avg1}) into the invader--resident comparison (\eqref{inv res comparison}) and grouping like-terms:

\begin{equation} \label{MCT full}
\begin{aligned}
    \overline{r}_i \approx & \underbrace{\alpha_i^{(1)} (\overline{E_i} - E_{i}^{*}) + \frac{1}{2} \alpha_i^{(2)} \Var{}{E_i} +  \beta_i^{(1)} C_{i}^{*}  - \sum \limits_{s \neq i}^S q_{is} \left((\overline{E_s} - E_{s}^{*}) + \frac{1}{2} \alpha_s^{(2)} \Var{}{E_s} +  \beta_s^{(1)} C_{s}^{*} \right) }_{r'_{i}, \text{Density-independent effects}}  \\ 
    + & \underbrace{\beta_i^{(1)} \overline{C_i} - \sum \limits_{s \neq i}^S q_{is}  \beta_s^{(1)} \overline{C_s} }_{\Delta \rho_i, \text{Linear density-dependent effects}} \\
    + & \underbrace{\frac{1}{2}\left[ \beta_i^{(2)} \Var{}{C_i}  - \sum \limits_{s \neq i}^S q_{is}   \beta_s^{(2)} \Var{}{C_s} \right]}_{\Delta N_i, \text{Relative nonlinearity}} \\
    + & \underbrace{\zeta_i  \Cov{}{E_i}{C_i}  - \sum \limits_{s \neq i}^S q_{is}  \zeta_s  \Cov{}{E_s}{C_s}.}_{\Delta I_i,  \text{The storage effect}}
\end{aligned}
\end{equation}

The new symbols ($r'_i$, $\Delta \rho_i$, $\Delta N_i$, and $\Delta I_i$) denote coexistence mechanisms.

One peculiar aspect of \eqref{MCT full} is that $r'_i$ contains $\beta_j^{(1)} C_{j}^{*}$ terms; $r'_i$ is the only coexistence mechanisms that contains multiple kinds of Taylor series terms. This quirk is related to the scaling factors. With the $\beta_j^{(1)} C_{j}^{*}$ terms shunted to  $r'_i$, the scaling factors can be used to cancel $\Delta \rho_i$ when there are more residents than limiting factors. Eliminating $\Delta \rho_i$ serves a definite role: it simplifies the invasion growth rate partition; allows us to not make the small-noise assumptions $C_j - C_j^* = \mathrm{O}(\sigma)$ and $\overline{C_j} - C_j^* = \mathrm{O}(\sigma^2)$, which are otherwise required; and highlights the role of fluctuation-dependent mechanisms by showing that not all species can be supported by classical mechanisms like resource partitioning (which are captured by $\Delta \rho_i$). However, we have argued (\citealp{johnson2022methodsb}) that empirical applications of MCT should keep the $\beta_j^{(1)} C_{j}^{*}$ terms in $\Delta \rho_i$, and should not use scaling factors.

Our exposition of MCT does not utilize the standard parameters (i.e., $\mathcal{E}_j$ and $\mathcal{C}_j$, see Eq. 6--9 in \citealp{Chesson1994}) since they impose an additional layer of potentially confusing abstraction, and because they lead to coexistence mechanisms that are quantitatively identical in the limit of small noise. That it not to say that the standard parameters are not useful --- they can be used to define coexistence mechanisms that sum exactly to the invasion growth (see \citealp{Chesson2019}; \citealp{Ellner2019}).

\end{enumerate}

The mathematical definition of the storage effect is

\begin{equation} \label{SE math def2}
    \Delta I_i = \zeta_i \;  \Cov{}{E_i}{C_i} - \sum \limits_{s \neq i}^S q_{is} \; \zeta_j \; \Cov{}{E_s}{C_s}.
\end{equation}

When ecologists talk colloquially about a storage effect, they are typically talking about a positive $\Delta I_i$ that is mediated through competition. However, the storage effect can also be negative, and/or mediated through apparent competition. In the case of a negative storage effect, there is a tendency for rarity to cause lower per capita growth rates. Therefore, negative storage effects can mediate a stochastic priority effect (\citealp{chesson1988interactions}; \citealp{schreiber2021positively}).

Coexistence mechanisms are often divided by the invader's sensitivity to competition, which we may operationalize here as $\abs{\beta_i^{(1)}}$. The rationale here is that $\abs{\beta_i^{(1)}}$ can be interpreted as the speed of population dynamics (at least in the lottery model and annual plant model; sensu \citealp{Chesson1994}), so dividing by it enables a better comparison of species with slow and fast life-cycles (\citealp{Chesson2018}). Note that this type of scaling is distinct from the aforementioned $q_{is}$ scaling factors.

Scaled coexistence mechanisms are sometimes averaged over species (see \citealp{chesson2003quantifying}, \citealp{barabas2018chesson}), either to make comparisons between communities or to quantify how a mechanism affects species \textit{in general}. The \textit{community-average storage effect} is defined as

\begin{equation} \label{community average SE}
   \overline{ \left(\frac{\Delta I}{\abs{\beta^{(1)}}} \right)} = \frac{1}{S} \sum_{i = 1}^{S} \frac{\Delta I_i}{\abs{\beta_i^{(1)}}}.
\end{equation}

\subsection{The ingredient-list definition}
\label{The ingredient list definition2}

The storage effect depends on three ingredients:

\begin{enumerate}
    \item species-specific responses to the environment,
    
    \item a non-zero interaction effect with respect to fluctuations in the environment and competition (also known as \textit{nonadditivity}), and
    
    \item covariance between environment and competition (\textit{$EC$ covariance} for short).
\end{enumerate}

We say that the storage effect "depends on three ingredients" (rather than "requires the ingredients"), because the ingredients' statuses as necessary and sufficient conditions are complex and context-dependent. On one hand, ingredients 2 \& 3 are necessary for the storage effect in the sense that $\Delta I_i$ will be zero if $\zeta_j = 0$ and $\Cov{}{E_j}{C_j} = 0$ for all $j$. Conditioned on the assumption of symmetric-species (see Appendix \ref{Symmetric species} below), all three ingredients are individually necessary and jointly sufficient for a non-zero $\Delta I_i$. On the other hand, ingredients 2 \& 3 are not necessary in the sense that one can construct examples where $\Delta I_i$ is positive despite some species (even the invader) having $\zeta_j \neq 0$ and $\Cov{}{E_j}{C_j} \neq 0$. To see why ingredient 1 is not necessary, consider the following non-symmetric scenario in which all species respond to the environment identically, but only species $k$ has a non-zero interaction effect (i.e., $\zeta_k \neq 0$). In this scenario, the condition \# 1 (species-specific responses to the environment) is not satisfied, yet species $k$ may nonetheless have a non-zero storage effect. In \ref{Quantifying the storage effect in the lottery model}, we show that the ingredients are not sufficient for the storage effect: when all ingredients are present in the lottery model, the storage effect can be zero if different species have different adult survival probabilities.

Our discussion of necessary and sufficient conditions does not imply that the ingredients are unimportant, nor that the ingredient-list definition has little value. For one, very few high-level features of the world have necessary and sufficient properties. In everyday life and in ecology, concepts are fuzzy, being held together by an open-ended set of correlated properties; \citet{wittgenstein1968philosophical} famously called these \textit{family resemblance concepts}. Secondly, the ingredient-list definition, when combined with math (\ref{Symmetric species}) and concrete examples (\ref{Species-specific responses to the environment}--\ref{Covariance between environment and competition}), is key to understanding the storage effect. Our discussion in the previous paragraph does, however, justify the usage of the "ingredient-list" metaphor --- the ingredients of a dish are not the dish itself; preparation also matters. 

\subsubsection{Symmetric species}
\label{Symmetric species}

To see how the simultaneous presence of all ingredients promotes coexistence \textit{in general}, we will consider the case of symmetric species.  Responses to the environment are temporally uncorrelated, but are correlated between species. Specifically, covariances are given by a symmetric covariance matrix, with $\Cov{}{E_j}{E_j} = \sigma^2$ and $\Cov{}{E_j}{E_k} = \rho \sigma^2$ when $j \neq k$. Otherwise, species are assumed to be demographically equivalent. 

At first, we will consider the case of two species: one resident and one invader. Suppose that both species share a competition parameter $C$ (as in the lottery model and annual plant plant model (\citealp{Chesson1994}, Section 5) that can be written as a smooth function $h$ of the resident's density and environmental response, i.e., $C = h(E_s, n_s)$. The competition parameter can then be written as a first-order Taylor series: $C = C^* + \frac{\partial h(E_s^*, n_s^*)}{\partial E_s}(E_s - E_s^*) + \frac{\partial h(E_s^*, n_s^*)}{\partial n_s}(n_s - n_s^*) + \mathcal{O}(\sigma^2)$. Plugging this Taylor series into the $EC$ covariance, we get

\begin{equation}
    \begin{aligned}
     \Cov{}{E_j}{C} & = \Cov{}{E_j}{C^* + \frac{\partial h(E_s^*, n_s^*)}{\partial E_s}(E_s - E_s^*) + \frac{\partial h(E_s^*, n_s^*)}{\partial n_s}(n_s - n_s^*)} + \mathcal{O}(\sigma^3) \\
     & = \Cov{}{E_j}{\frac{\partial h(E_s^*, n_s^*)}{\partial E_s} E_s + \frac{\partial h(E_s^*, n_s^*)}{\partial n_s} n_s } + \mathcal{O}(\sigma^3) \quad \left[\text{since } \Cov{}{a+X}{b+Y} = \Cov{}{X}{Y} \right] \\
     & = \Cov{}{E_j}{\frac{\partial h(E_s^*, n_s^*)}{\partial E_s} E_s} + \mathcal{O}(\sigma^3) \quad \left[\text{since $E_j$ is temporally uncorrelated} \right] \\
     & = \frac{\partial h(E_s^*, n_s^*)}{\partial E_s} \Cov{}{E_j}{E_s} + \mathcal{O}(\sigma^3) \quad \left[\text{since } \Cov{}{cX}{dY} = cd\Cov{}{X}{Y} \right].
    \end{aligned}
\end{equation}

With the symmetric covariance matrix and the simplified notation $\theta =  \frac{\partial h(E_s^*, n_s^*)}{\partial E_s}$,  the covariance approximation becomes

\begin{equation}
    \Cov{}{E_j}{E_s} \approx 
    \begin{cases}
    \theta \sigma^2 & j = s \\  
    \theta \rho \sigma^2 & j \neq s \\  
    \end{cases}
\end{equation}

Because species are demographically equivalent (with the exception of partially uncorrelated responses to the environment), they have quantitatively identical Taylor series coefficients, i.e., $\zeta_j = \zeta$, and $\beta_j^{(1)} = \beta^{(1)}.$ The scaling factors are $q_{is} = \frac{\beta_i^{(1)}}{\beta_s^{(1)}} \frac{\partial C_i}{\partial C_s}$ = 1. The storage effects takes the strikingly simple form, 

\begin{equation}
    \Delta I \approx -\zeta (1-\rho) \theta \sigma^2.
\end{equation}

All three ingredients are represented in this expression. \textit{species-specific responses to the environment} are captured by $1-\rho$, a measure of environmental niche overlap. The interaction effect, $\zeta$, is generally negative in models of resource competition, thus making the storage effect positive. The covariance between environment and competition is captured by $\theta \sigma^2$ and $\rho \theta \sigma^2$ for the resident and invader respectively.

Generalizing to the multi-species, involves redefining the competition-generating function $h$ so that competition is determined all residents' environmental responses, $\myvect{E}$, and population densities, $\myvect{n}$. The parameter $\theta$ is also redefined: $\theta = \frac{\partial h(\myvect{E^*}, \myvect{n^*})}{\partial E_s}$, where $s \neq i$. 

The invader's $EC$ covariance is 
\begin{equation}
\begin{aligned}
    \Cov{}{E_i}{C} & \approx \sum_{s \neq i}^S \frac{\partial h(\myvect{E^*}, \myvect{n^*})}{\partial E_s} \Cov{}{E_i}{E_s} \\
    & \approx \theta \sigma^2 \rho (S-1) ,
\end{aligned}
\end{equation}

whereas the residents' $EC$ covariance is

The invader's $EC$ covariance is 
\begin{equation}
\begin{aligned}
    \Cov{}{E_s}{C} & \approx \sum_{k \neq i}^S \frac{\partial h(\myvect{E^*}, \myvect{n^*})}{\partial E_k} \Cov{}{E_s}{E_k} \\
    & \approx \theta \sigma^2 \left( \rho (S-2) + 1 \right).
\end{aligned}
\end{equation}

Again, the storage effect is $\Delta I \approx -\zeta (1-\rho) \theta \sigma^2$. However, there is one difference between the multi-resident case and the single-resident case. In the multi-resident case $\theta$ tends to be inversely proportional to the number of residents --- multiple residents affect the competition parameter, so a single resident's environmental response has a smaller effect on the competition parameter.

For example, the competition parameter in the lottery model is $C = \log( \frac{\sum \limits_{k = 1}^{S} \exp(E_k) n_{k} }{\sum \limits_{k = 1}^{S} \delta_k n_{k}})$, where $\delta_k$ is the morality probability of adult fish (see Appendix \ref{Quantifying the storage effect in the lottery model} for details). With this choice of $C$, along with the symmetric-species assumption, we find that $\theta = 1/(S-1)$. The coefficient $\theta$ generically scales with $1/(S-1)$ whenever competition can be written as the logarithm of a linear combination of environmental responses. While the storage effect can theoretically support an arbitrary number of species with a single regulating factor, the storage effect becomes weaker as communities become more speciose. As a consequence, coexistence becomes less robust --- small deviations from the symmetric-species case are likely to result in extirpations --- once again demonstrating the "\ldots impossibility of coexistence of infinitely many strategies" (\citealp{Gyllenberg2005}).

\subsubsection{Ingredient \#1: Species-specific responses to the environment}
\label{Species-specific responses to the environment}

The function of ingredient \#1 is rather obvious: to establish the presence of niche differences, which are necessary for coexistence via any mechanism (\citealp{Gause1934}; \citealp{chesson1991need}). In the absence of \textit{species-specific responses to the environment}, there would be no rare-species advantage. In terms of the lottery model,  a good (bad) year for the blue species would automatically be a good (bad) year for the common red species, such that both species would always experience the same level of competition. Interestingly, ingredient \#1 is conceptually intuitive but mathematically obscure --- ingredient \#1 manifests mathematically in the differential magnitudes of the invader and residents' $EC$ covariances, but seeing this clearly requires simplifying assumptions (as in the case of \textit{symmetric-species}, Appendix \ref{Symmetric species}).

\subsubsection{Ingredient \#2: An interaction effect between environment and competition}
\label{An interaction effect between environment and competition}

When a red fish experiences a hot year, it is not merely the case that the negative effects of high competition offset the positive effects of a good environment. Rather, the environment and competition act synergistically to reduce per capita growth rates further. This synergy is the \textit{interaction effect}, akin to an interaction effect in multiple regression. In fact, the coefficient $\zeta_j$ in the mathematical definition of the storage effect (\eqref{SE math def2}) \textit{is} the interaction effect of a multiple regression, in the limit of small environmental noise, where $E_j - E_j^*$ and $C_j - C_j^*$ are predictors. The causal interpretation of an interaction effect in multiple regression is that the level of one predictor modulates another predictor's effect on the response variable (\citealp{gelman2007data}), which is why the simple interpretation of the storage effect (see the main text, Section: \ref{An okay interpretation of the storage effect}) states that "high competition undermines the positive effects of a good environment". Put yet another way, high competition means that the population is less sensitive to changes in the environment.

In our exposition thus far, we have described a negative interaction effect. However, both negative or positive interaction effect can lead to either a positive or negative storage effect. In the jargon of Modern Coexistence Theory, a negative interaction effect (i.e., $\zeta_j < 0$) is called \textit{subadditivity} or \textit{buffering} (\citealp{Chesson1994}). The term \textit{subadditive} comes from the fact that the joint effects of environment and competition are less than than the sum of their parts. The term \textit{buffering} comes from the fact that the doubly deleterious effect of a poor environment and high competition is somewhat abated: the term $\zeta_j (E_j - E_j^*)(C_j-C_j^*)$ is positive when $\zeta_j < 0$, $(E_j - E_j^*) < 0$, and $(C_j-C_j^*) >0$. A positive interaction effect (i.e., $\zeta_j >0$) is synonymous with \textit{superadditivity} or \textit{amplifying} (\citealp{Chesson1989}). More generally, both positive and negative interaction effects are referred to as \textit{nonadditivity}. The storage effect is generally positive in systems with subadditivity and positive $EC$ covariances, \textit{or} systems with superadditivity and negative $EC$ covariances. Conversely, the storage effect is generally negative in systems with subadditivity and negative $EC$ covariances, \textit{or} systems with superadditivity and positive $EC$ covariances.

At a high level of abstraction, the interaction effect can be thought of combining the environment and competition into a large number of density-dependent factors. Coexistence requires negative feedback loops where species demonstrate some degree of specialization on density-dependent factors (\citealp{meszena2006competitive}), but some communities do not have enough density-dependent factors to be specialized upon (\citealp{hutchinson1959homage}; \citealp{hutchinson1961paradox};  but see \citealp{levin1970community}; \citealp{haigh1972can}; \citealp{abrams1988should}). On the other hand, species may readily specialize on different environmental states, but environmental variation alone cannot promote coexistence (\citealp{chesson1997roles}). The interaction effect combines the competition parameter with the environmental parameter to get the best of both worlds: the density-dependent factors (implicit in the competition parameter) provide the negative feedback while the species-specific environmental parameter provides the specialization.

But what is an interaction effect in more concrete terms? In the literature, a negative interaction effect has primarily been associated with differential sensitivities of different life-stages. \citet{chesson1988community} write, "... iteroparous plant and sessile marine organisms, can buffer by participating in reproduction over a number of years... Semelparous species can experience these buffering effects if the offspring of an individual mature over a range of years..." More generally, a negative interaction effect may arise from other forms of population structure: dormancy (\citealp{caceres1997temporal}; \citealp{ellner1987alternate}), phenotypic variation (\citealp{chesson2000mechanisms}), or spatial variation (\citealp{chesson2000general}). In studies of the population genetic storage effect (which promotes allelic diversity), negative interaction effects can be produced by heterozygotes (\citealp{Dempster1955}; \citealp{Haldane1963}), sex-linked alleles (\citealp{reinhold2000maintenance}), epistasis (\citealp{gulisija2016phenotypic}), and maternal effects (\citealp{yamamichi2017roles}). 

When fecundity fluctuates and an adult life-stage is insensitive to the environment and competition, then adult survival will lead to a negative interaction effect --- adults are simply not affected by the joint occurrence of a poor environment and high competition. If, on the other hand, adult survival fluctuates, then adults are disproportionately hurt by a poor environment (i.e., low adult survival) and high competition (\citealp{chesson1988interactions}). Of course, an interaction effect does not require population structure. Interaction effects results from per capita growth rates with multiplicative functional forms; see the phytoplankton model (\eqref{phyto}) or our empirically-driven annual plant model (\eqref{fruit}) from the main text. The ecological interpretation of an interaction effect (or lack thereof) must be determined on a model-by-model basis, either with mathematical analysis or analogy with previously-studied models.  

\subsubsection{Ingredient \#3: Covariance between environment and competition}
\label{Covariance between environment and competition}

The final ingredient, covariance between environment and competition, is immediately evident in the mathematical definition of the storage effect (\eqref{SE math def2}). In more biological terms, the covariance captures the causal relationship between environment and competition. The most obvious way in which a good environment causes high competition (and vice-versa) is through intergenerational population growth: a good environment produces a larger population, and a larger population usually corresponds to higher competition. However, temporal autocorrelation in the environment is required for $EC$ covariance via intergenerational population growth  (\citealp{li2016effects}; \citealp{letten2018species}; \citealp{Ellner2019}; \citealp{schreiber2021positively}); the past environment determines the present competition, but the $EC$ covariation involves the current environment and current competition, so the current environment must resemble the past environment. The storage effect also arises when species have phenology differences in periodic environments (\citealp{loreau1989coexistence}; \citealp{loreau1992time}; \citealp{klausmeier2010successional}), since a periodic environment is just a special case of a temporally autocorrelated environment.  In a stage-structured model, a good environment can lead to high competition within a single time-step (\citealp{chesson1988community}). Consider the lottery model: a good environment (i.e., high per capita fecundity) at the spawning stage leads to high competition (i.e., many larvae per territory) at the recruitment stage. Note here that there is still temporal autocorrelation in the sense that the larvae carry the effects of the environment through time.

Although the archetypical storage effect is mediated through resource competition, the storage effect may also be mediated through apparent competition; the parameter $C_j$ may be generally understood as the effects of all density-dependent factors. When the storage effect is mediated through resource competition, the $EC$ covariance is generically positive, though it may be negative when the environment is negatively autocorrelated (\citealp{schreiber2021positively}). 

When the storage effect is mediated through apparent competition, a negative, positive or zero-valued covariance is possible. For the sake of the current discussion, assume that the competition parameter is the density of the shared predator $P$, times the predator's functional response $f_j(N_j)$, all divided by prey density i.e., $C_j = P * (f_j(N_j) / N_j)$. \citet{stump2017optimally} analyzed a variant of the annual plant model and found that a type 2 functional response leads to a negative $EC$ covariance: good environments lead to a large number of seeds, which satiate predators, thus lowering the per-seed predation pressure. Kuang and Chesson (\cite*{kuang2010interacting}, \cite*{chesson2010storage}) found that the a type 3 functional response (i.e., frequency-dependent predation) leads to a positive $EC$ covariance: good environments lead to a large number of seeds, which are then preferentially consumed. In the previous two examples, the predator demonstrates a fast behavioral response to changes in prey density. If the predator demonstrates only a numerical response to prey density (corresponding to a type 1 functional response) and the environment is temporally autocorrelated, then the covariance will be positive. If instead the environment is uncorrelated through time, the storage effect due to the predation is not possible (\citealp{Kuang2009CoexistenceEffect}): the environment changes before it appreciably affects predation pressure, and therefore does not produce the necessary covariance. 

\section{The storage effect in the lottery model}
\label{Quantifying the storage effect in the lottery model}

The lottery model with pure temporal variation (\textit{sensu} \citealp{Chesson1994}) is written as
\begin{equation} \label{lottery2}
   n_{j}(t+1) =  n_{j}(t) \left[ s_j + \eta_{j}(t)   \left(\rule{0cm}{1.25cm}\right. \frac{ \sum \limits_{k = 1}^{S} (1 - s_k) n_{k}(t)}{\sum \limits_{k = 1}^{S} \eta_{k}(t) n_{k}(t)} \left.\rule{0cm}{1.25cm}\right) \right], \quad j = (1, 2, \ldots, S).
\end{equation}
Note that we do not need extra equations to track larvae because they die if they are not recruited. In order to fit the lottery model into the mold of Modern Coexistence Theory, we must define the environmental parameter and the competition parameter. Selecting $E_j = \log(\eta_j)$ and $C = \log( \frac{\sum \limits_{k = 1}^{S} \eta_{k} n_{k} }{\sum \limits_{k = 1}^{S} (1-s_k) n_{k}})$, the effective per capita growth rate takes the form 

\begin{equation} \label{lotteryEC}
   r_j(E_k,C_j) = \log(s_j + \exp{E_j - C})
\end{equation}

With the log-scale specifications of $E_j$ and $C$, all Taylor series coefficients can be expressed purely as a function of $s_j$, which leads to tidy expressions of the coexistence mechanisms. If one uses the non-log-scale specification, the results are qualitatively identical.

Next, we choose the equilibrium parameters $E_j^*$ and $C_j^*$. The shared equilibrium level of competition is defined as the temporal average of competition experienced by the invader, averaged over all species acting as the invader: $C^* = \frac{1}{S} \sum_{i=1}^{S} \overline{C_{i}}$. With this choice made, the equilibrium environmental parameters are fixed at $E_j^* = \log(1 - s_j) + C^*$.

With the equilibrium parameters defined, the partial derivatives of $g_j$ may be computed:

\begin{equation}  
\begin{aligned}
& \alpha_j^{(1)}  = \pdv{g_j(E_j^*, C_j^*)}{E_j} = 1-s_j\\
 & \beta_j^{(1)} = \pdv{g_j(E_j^*, C_j^*)}{C_j} = -(1-s_j)\\ 
 & \alpha_j^{(2)} = \pdv[2]{g_j(E_j^*, C_j^*)}{E_j} = s_j (1-s_j)\\ 
 & \beta_j^{(2)} = \pdv[2]{g_j(E_j^*, C_j^*)}{C_j} = s_j (1-s_j)\\ 
 & \zeta_j = \pdv{g(E_j^*, C_j^*)}{E_j}{C_j} = -s_j(1-s_j).\\
\end{aligned}
\end{equation}

The general definition of the scaling factors is $q_{is} = \frac{\beta_i^{(1)}}{\beta_s^{(1)}} \frac{\partial C_i}{\partial C_s}$. In the case where species share a single competition parameter, this reduces to $q_{is} = \frac{\beta_i^{(1)}}{\beta_s^{(1)}}$.

Under the standard small-noise assumptions of Modern Coexistence Theory, the competition parameter can be approximated by a function of residents' environmental responses: $(C - C^*) \approx \sum_{s \neq i}^S \frac{\partial h(\myvect{E^*}, \myvect{n^*})}{\partial E_s} (E_s - E_s^*)$. In the lottery model with only one invader and one resident, there is a one-to-one conversion between environment and competition: $\frac{\partial  h(E_s^*, n_s^*)}{\partial E_s} = 1$. 

In a two-species system with species $1$ as the invader and species $2$ as the resident, the covariance between species $1$'s environment response and competition is $\Cov{}{E_1}{C} = \Cov{}{E_1}{E_2} = \rho \sigma^2$, where $\rho$ is the correlation between the two species' responses to the environment, and $\sigma^2$ is the variance of environmental responses (shared by both species for maximum simplicity).

Utilizing the general mathematical expression of the storage effect, we find that the storage effect of species $1$ is 
\begin{equation} \label{SE simple1}
    \Delta I_1 \approx  \sigma^2 \left( s_2(1-s_1) - \rho \; s_1(1-s_1) \right).
\end{equation}
Setting $\Delta I_1$ equal to zero and solving for $s_2$, we find that $s_2 = \rho s_1$, which has a solution when $\rho$ is positive. This shows that the storage effect can be zero even when there are species-specific responses to the environment (i.e., $\rho < 1$), subadditivity (i.e., $s_j > 0$), and covariance between environment and competition (i.e., $\sigma^2 \neq 0$), as claimed in the main text (Section: \ref{The ingredient list definition}).

Coexistence mechanisms are often divided by species' sensitivity to competition, here operationalized as $\abs{\beta_i^{(1)}}$. This scaled version of species 1's storage effect is 

\begin{equation} \label{SE simple2}
    \frac{\Delta I_1}{(1-s_1)} \approx  \sigma^2 \left( s_2 - \rho \; s_1 \right).
\end{equation}

To obtain the effects of all variability on the invasion growth rate, we first introduce the notation $\Delta E_i'$ for the nonlinear effects of the environment:
\begin{equation}
    \Delta E_i' = \frac{1}{2} \left[ \beta_i^{(2)} \Var{}{E_i} - \sum_{s \neq i} q_{is} \beta_s^{(2)} \Var{}{E_s} \right].
\end{equation}
Following the definitions of coexistence mechanisms (\eqref{MCT full}), the scaled sum of the scaled storage effect, relative nonlinearity, and the nonlinear effects of the environment for species $1$ is 
\begin{equation} \label{SE simple3}
    \frac{\Delta I_1 + \Delta N_1 + \Delta E_1'}{(1-s_1)} \approx \sigma^2 \, s_1(1-\rho), 
\end{equation}
which clearly increases with invader survival, as claimed in Appendix \ref{An alternative definition of the storage effect}.

In the two-species lottery model, the scaled community-average storage effect (\eqref{community average SE}) is
\begin{equation} \label{community average SE lottery}
   \overline{ \left(\frac{\Delta I}{1-s} \right)} = \frac{\sigma^2}{2}(s_1 + s_2)(1-\rho),
\end{equation}
which demonstrates that at the community-level, survival increases the (community-average) storage effect, as claimed in the main text (Section: \ref{Discussion: Rethinking storage}).

To see that the adult survival is not a sufficient condition for the storage effect, consider a modified lottery model where adult survival probability fluctuates. If we identify the environmental parameter as $E_j = s_j$ and the equilibrium parameter as $E_j = \overline{s_j}$ (the temporal mean), the per capita growth rate function can be written as $g_j(E_j, C_j) = \log(E_j + \eta_j \exp{-C_j})$ (compare with \eqref{lotteryEC}), leading to the interaction effect, $\gamma_j = 1-\overline{s_j}$. The storage effect goes to zero as adults become more robust.

\section{Bet-hedging and the storage effect}
\label{bet_hedging}

In discrete time models, the average per capita growth rate is the temporal average of the logged finite rate of increase: $\overline{\log(\lambda_j)}$. If we assume that $\lambda_j - 1 = \mathrm{O}(\sigma)$, a foundational assumption in Modern Coexistence Theory (\citealp{Chesson1994}, \citealp{chesson2000general}), the average per capita growth rate can be approximated with a Taylor series expansion about 1:

\begin{equation} \label{loglambda}
    \overline{\log(\lambda_j)} = \overline{\lambda_j} - 1 - \frac{1}{2} \overline{\left(\lambda_j - 1 \right)^2} + \mathrm{O}(\sigma^3).
\end{equation}

If we make the additional (but also foundational) assumption that $\overline{\lambda_j} - 1 = \mathrm{O}(\sigma^2)$, then \eqref{loglambda} can be re-written as 

\begin{equation} \label{loglambda2}
    \overline{\log(\lambda_j)} = \overline{\lambda_j} - 1 - \frac{1}{2} \Var{}{\lambda_j} + \mathrm{O}(\sigma^3).
\end{equation}

Now we see that species can improve their average per capita growth rate, (also known as the stochastic growth rate), by either increasing their \textit{mean fitness}, $\overline{\lambda_j}$, or by decreasing their \textit{temporal fitness variation}, $\Var{}{\lambda_j}$. Evolutionary bet-hedging, a type of risk aversion, occurs when species evolve some adaptation that reduces fitness variation at the cost of also reducing average fitness.  

The finite rate of increase can be expressed as a function $g'_j$ of $E_j$ and $C_j$. In analogy with \eqref{taylor_decomp}, we can expand $\lambda_j$ about the equilibrium parameters, selected so that $g_j^\prime(E_j^*, C_j^*)=1$, resulting in the approximation
\begin{equation} \label{taylor_decomp2}
\begin{aligned}
\lambda_{j}
 \approx \; & \alpha_j^{\prime(1)} (E_j - E_{j}^{*}) + \beta_j^{\prime(1)} (C_j - C_{j}^{*}) \\ & + 
\frac{1}{2} \alpha_j^{\prime(2)} (E_j - E_{j}^{*})^{2} + \frac{1}{2} \beta_j^{\prime(2)} (C_j - C_{j}^{*})^{2} + 
\zeta_j^\prime  (E_j - E_{j}^{*})   (C_j - C_{j}^{*}),
\end{aligned}
\end{equation}
with the following Taylor series coefficients:
\begin{equation}  
\begin{aligned}
 \alpha_j^{\prime(1)} = \pdv{g_j^\prime\scriptstyle{(E_j^*, C_j^*)}}{E_j},  \quad
 \beta_j^{\prime(1)} = \pdv{g_j^\prime\scriptstyle{(E_j^*, C_j^*)}}{C_j},  \quad
 \alpha_j^{\prime(2)} = \pdv[2]{g_j^\prime\scriptstyle{(E_j^*, C_j^*)}}{E_j},  \quad
 \beta_j^{\prime(2)} = \pdv[2]{g_j^\prime\scriptstyle{(E_j^*, C_j^*)}}{C_j,}  \quad
 \zeta_j\prime = \pdv{g\scriptstyle{(E_j^*, C_j^*)}}{E_j}{C_j}.
\end{aligned}
\end{equation}
Note the "prime" superscript, which indicates that the approximation and coefficients are not identical to \eqref{taylor_decomp} \& \eqref{taylor_coef}, where the effective per capita growth rate (not the finite rate of increase) was approximated.

Now we can substitute the approximation of $\lambda_j$ (\eqref{taylor_decomp2}) into the bet-hedging partition of $\overline{\log(\lambda_j)}$ (\eqref{loglambda2}). Performing this substitution, replacing the second-order terms with central moments (e.g., $(E_j - E_j^*)^2 = \Var{}{E_j} + \mathrm{O}(\sigma^3)$), and truncating at second-order, we obtain

\begin{equation}
\begin{array}[b]{ll}
    \overline{\log(\lambda_j)} \approx  & 
    \left.
    \begin{array}[t]{l}
       \alpha_j^{\prime(1)} (\overline{E_j} - E_{j}^{*}) + \beta_j^{\prime(1)} (\overline{C_j} - C_{j}^{*}) \\
         + \frac{1}{2} \alpha_j^{\prime(2)} \Var{}{E_j} + \frac{1}{2} \beta_j^{\prime(2)} \Var{}{C_j} + 
\zeta_j^\prime  \Cov{}{E_j}{C_j}
    \end{array} 
\right \}= \overline{\lambda_j} - 1 \\
   & \left.
    \begin{array}{l}
         - \frac{1}{2}\left(\alpha_j^{\prime(1)}\right)^2 \Var{}{E_j} - \frac{1}{2}\left(\beta_j^{\prime(1)}\right)^2 \Var{}{C_j} \\
         - \alpha_j^{\prime(1)} \beta_j^{\prime(1)} \Cov{}{E_j}{C_j}
    \end{array}
\right \}= \Var{}{\lambda_j}
\end{array}
\end{equation}

We can extract the storage effect by collecting all terms containing $\Cov{}{E_j}{C_j}$, and then performing the invader--resident comparison (\eqref{inv res comparison}). The storage effect is 

\begin{equation} \label{SE math def3}
    \Delta I_i = \left( \zeta_i^\prime - \alpha_i^{\prime(1)} \beta_i^{\prime(1)} \right) \;  \Cov{}{E_i}{C_i} - \sum \limits_{s \neq i}^S q_{is} \; \left( \zeta_s^\prime - \alpha_s^{\prime(1)} \beta_s^{\prime(1)} \right)  \; \Cov{}{E_s}{C_s}.
\end{equation}

The expression clearly shows that the storage effect affects both mean fitness and temporal fitness variation. We can split the storage effect into two parts, the storage effect mediated through mean fitness, and the storage effect mediated through fitness variation, respectively defined as

\begin{equation}
    \Delta I_i^{\{\text{mean}\}} = \zeta_i^\prime \;  \Cov{}{E_i}{C_i} - \sum \limits_{s \neq i}^S q_{is} \;  \zeta_s^\prime \; \Cov{}{E_s}{C_s}, \quad \text{and}
\end{equation}

\begin{equation}
    \Delta I_i^{\{\text{var}\}} = - \alpha_i^{\prime(1)} \beta_i^{\prime(1)} \;  \Cov{}{E_i}{C_i} + \sum \limits_{s \neq i}^S q_{is} \;  \alpha_i^{\prime(1)} \beta_i^{\prime(1)} \; \Cov{}{E_s}{C_s}.
\end{equation}

To see how each term \textit{generally} affects coexistence, we will make that same simplifying assumptions that we used in the case of \textit{symmetric species} (Appendix \ref{Symmetric species}). Species have a shared competition parameter that can be written as a function of population densities and environmental parameters; covariances between species' $E_j$ are given by a symmetric covariance matrix with $\sigma^2$ on the diagonal and $\rho \sigma^2$ on the off-diagonals; and species otherwise have identical demographic parameters, such that species have identical Taylor series coefficients (and we have no need for species-specific subscripts). 

Following the derivation in in Appendix \ref{Symmetric species}, we see that $\Cov{}{E_i}{C} = \theta \rho \sigma^2$ and $\Cov{}{E_s}{C} = \theta \sigma^2$; recall that $\theta$ is a positive constant that converts residents' environmental responses into competition. Now $\Delta I_i^{\{\text{mean}\}}$ and $\Delta I_i^{\{\text{var}\}}$ can be simplified:

\begin{equation}
    \Delta I_i^{\{\text{mean}\}} = - \zeta^\prime \left( 1- \rho \right) \theta \sigma^2,
\end{equation}

\begin{equation}
    \Delta I_i^{\{\text{var}\}} =  \alpha_i^{\prime(1)} \beta_i^{\prime(1)} \left( 1- \rho \right)
\end{equation}

\begin{figure} 
  \centering
      \includegraphics[scale = 0.6]{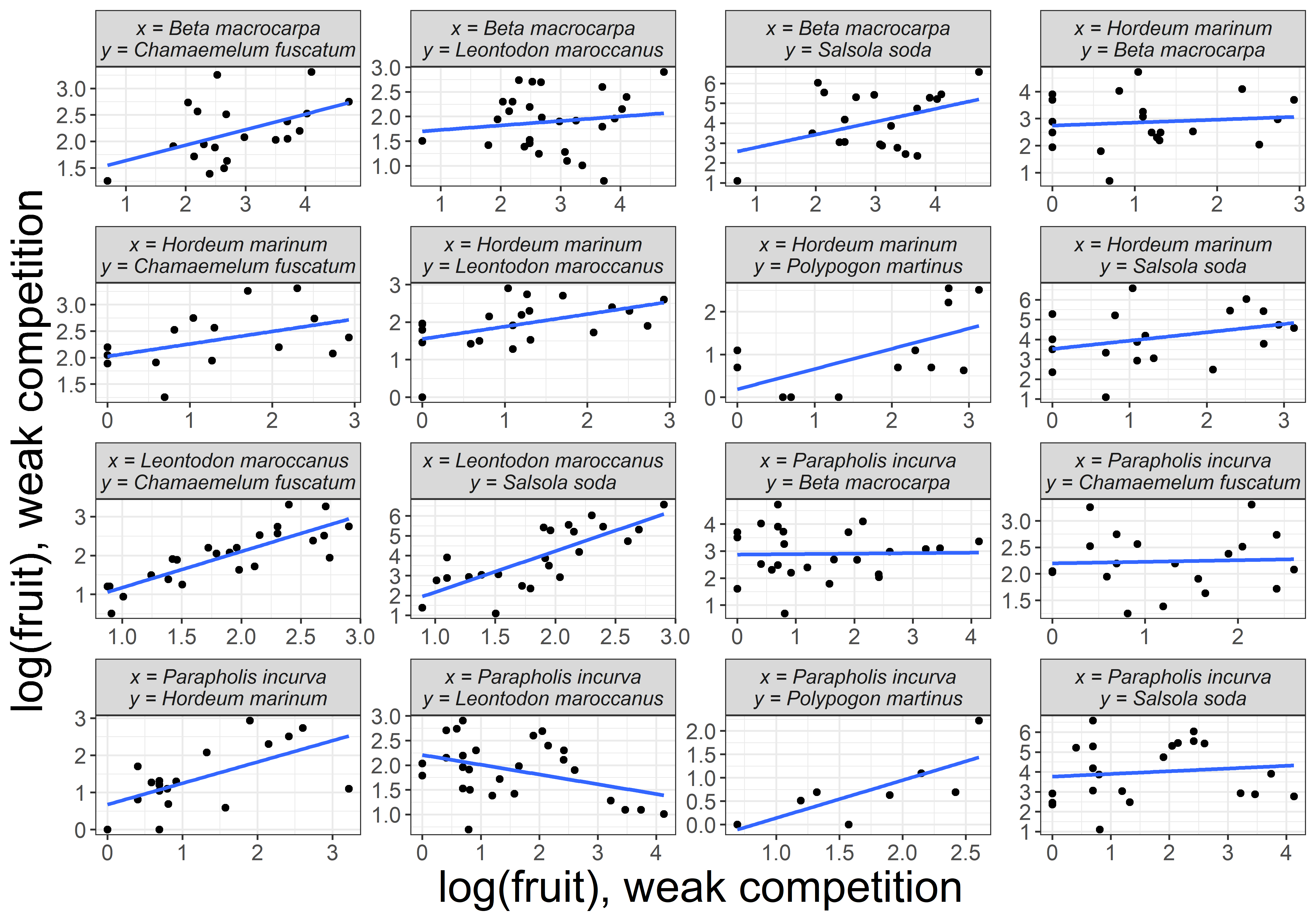}
  \caption{"Model-free" evidence of ingredient 1: species-specific responses to the environment. Pair plots show that species pairs tend to have weakly correlated responses to the environment. Each point represents the logarithm of the average fruit produced by two species in the same growing season and plot (spatial and temporal proximity are shorthand for a similar environment). The data has been subsetted to individuals experiencing weak competition (less than the 15th percentile of $\sum_k N_k$), allowing us to isolate the effects of the environment.  \label{pairs}}
\end{figure}

In general, $I_i^{\{\text{mean}\}} > 0$ and $I_i^{\{\text{var}\}} < 0$. To see this, note that $\alpha_i^{\prime(1)} > 0$ (by convention in MCT, a large environmental response is good for population growth), $\beta_i^{\prime(1)} < 0$ (i.e., competition is bad for population growth), $0 < \rho \leq 1$, and $\zeta^\prime < 0$. The correlation $\rho$ tends to be positive (see Section \ref{A critique of the conventional interpretation of the storage effect}; also Fig. \ref{pairs}) though this is not guaranteed. Similarly, $\zeta^\prime < 0$ tends to occur in models of resource competition. For example, the lottery model exhibits a negative interaction effect on the natural scale, even when there is no adult survival (\citealp{chesson2000general}).

The usual signs of the storage effect terms imply the inequalities $I_i^{\{\text{mean}\}} > 0$ and $I_i^{\{\text{var}\}} < 0$, which show that the storage effect exacerbates the negative effects of temporal fitness variation, and compensates by increasing mean fitness. Thus, the storage effect promotes coexistence via the inverse of bet-hedging.

\section{Alternative definitions of the storage effect}
\label{An alternative definition of the storage effect}

An alternative definition of the storage effect could use an \textit{invader--invader comparison} (\citealp{johnson2022methods}): instead of comparing an invader to residents (as in \eqref{inv res comparison}), we could compare a single focal species to itself at high versus low density. Using this alternative definition of the storage effect in the two species lottery model, we would find that the storage effect increases with adult survival: when going from the high-density state to the low-density state, the $EC$ covariance decreases (assuming that there are species-specific responses to the environment) but the survival parameter --- and thus the interaction effect $\zeta_j$ --- is unchanged. 

There are problems with the invader--invader comparison that make it a poor method for defining coexistence mechanisms generally (\citealp{johnson2022methods}). It is not defined when the invasion growth rate is negative, since a community with the focal species at high density cannot be prepared for the purpose of the comparison. Additionally, the invader--invader comparison does not necessarily capture the notion of specialization/differentiation. However, even if these problems could be circumvented, the invader--invader comparison would not validate the conventional interpretation v2. The invader--invader comparison does not say that buffering helps rare species. Rather, the invader--invader comparison shows that buffering helps a rare species relative to said species at high density: it is as much about self-limitation when abundant as it is about recovery when rare.

One could argue that the storage effect should be defined as $\Delta N_i + \Delta I_i$, the sum of the contemporary relative nonlinearity and storage effect, at least in some cases. In the lottery model, the environment is the ultimate origin of variation in competition. Since $\Delta N_i$ captures the rare-species advantage due to variation in competition, the sum $\Delta N_i + \Delta I_i$ captures all of the coexistence-promoting effects, direct and indirect, of environmental variation. Indeed, early formulae defined the storage effect in this way (\citealp{warner1985coexistence}, Eq.10).

If the storage effect was to be redefined as $\Delta N_i + \Delta I_i$, increasing invader survival increases the invader's storage effect (at least in the lottery model), regardless of whether species have positively or negatively correlated responses to the environment. However, this definition of the storage effect does not isolate the effect of \textit{buffering}, and thus the conventional interpretation v2 does not apply.

Finally, we acknowledge that our empirical criticism of the conventional interpretation v2 (Section \ref{A critique of the conventional interpretation of the storage effect} in the main text) may be missing the point. An imprecise definition or interpretation can often serve a pedagogical purpose. One example comes from \citet{godfrey2009darwinian}: Many evolution textbooks state that there are three conditions for evolution by natural selection: phenotypic variation, heritability, and fitness differences. However, these conditions are not sufficient for evolution, since all conditions can be met, and yet allele frequencies remain stagnant in the face of stabilizing selection. Does the conventional interpretation v2 of the storage effect similarly fall into the category of imperfect-but-useful? By questioning the generality of the conventional interpretation v2, are we undermining its (implicitly pedagogical) purpose? Perhaps; buffering promotes persistence when species' responses to the environment are negatively correlated, and perhaps it is easier to think of niche differentiation (a necessary ingredient for coexistence via any mechanism) as a negative correlation between environmental responses, rather than positively but not perfectly correlated responses. 

\section{Model-fitting details} 
\label{model fitting details}

All code for model-fitting, analysis, and figure production is available at \url{https://github.com/ejohnson6767/storage_effect_critique}. Models were fit using the \textit{Stan} program's implementation of Hamiltonian Monte Carlo, running on 4 chains for 1000 iterations each. Diagnostic statistics ("R hat", the effective sample size, energy levels) were examined to ensure that all chains had converged to the same posterior distribution. All parameters were assigned weakly informative priors. The standard deviations of marginal posterior distributions were much smaller than the standard deviations of the corresponding priors, indicating that our prior choices had negligible influence.

Due to the large sample size and the complexity of the models (specifically the sheer number of competition coefficients), we utilized a supercomputing cluster along with \textit{Stan}'s capability for \textit{within-chain parallelization}. We have made our model-fit objects available on \textit{GitHub}, but readers hoping to replicate everything should be aware that fitting individual models with 64 cores can take up 2 hours.

\begin{figure} 
  \centering
      \includegraphics[scale = 0.6]{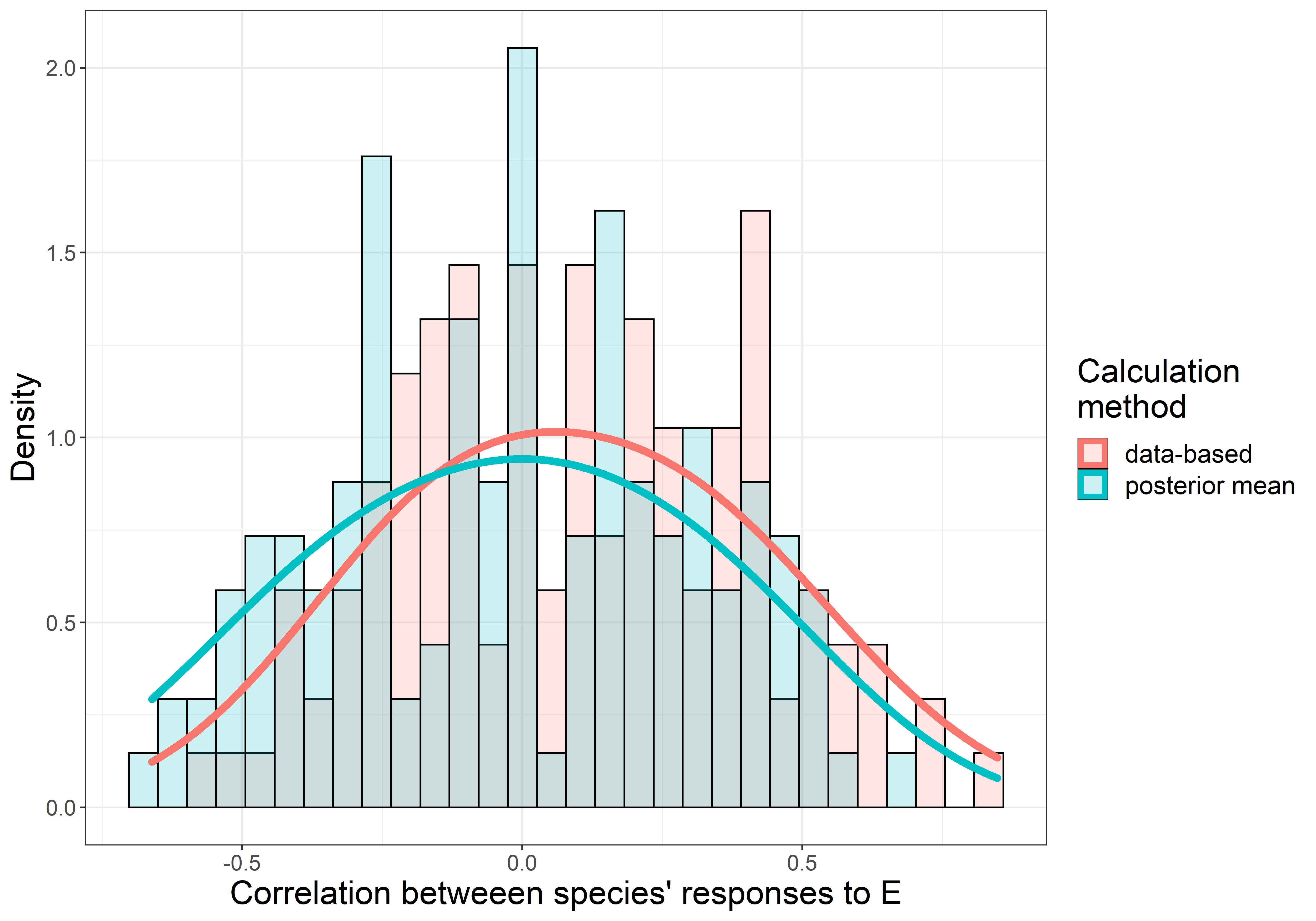}
  \caption{Distribution of pairwise correlations between species' responses the environment, across all combinations of species pairs. The "data-based" method calculates the correlation in species pair plots, using the methodology laid out in Figure \ref{pairs}. The "model-based" method calculates the posterior average of $\mathrm{Cor}\left(\beta_{1,j} M, \beta_{1,k} M \right)$, for each unique species pair. \label{response_hist}}
\end{figure}

\begin{figure} 
  \centering
      \includegraphics[scale = 0.6]{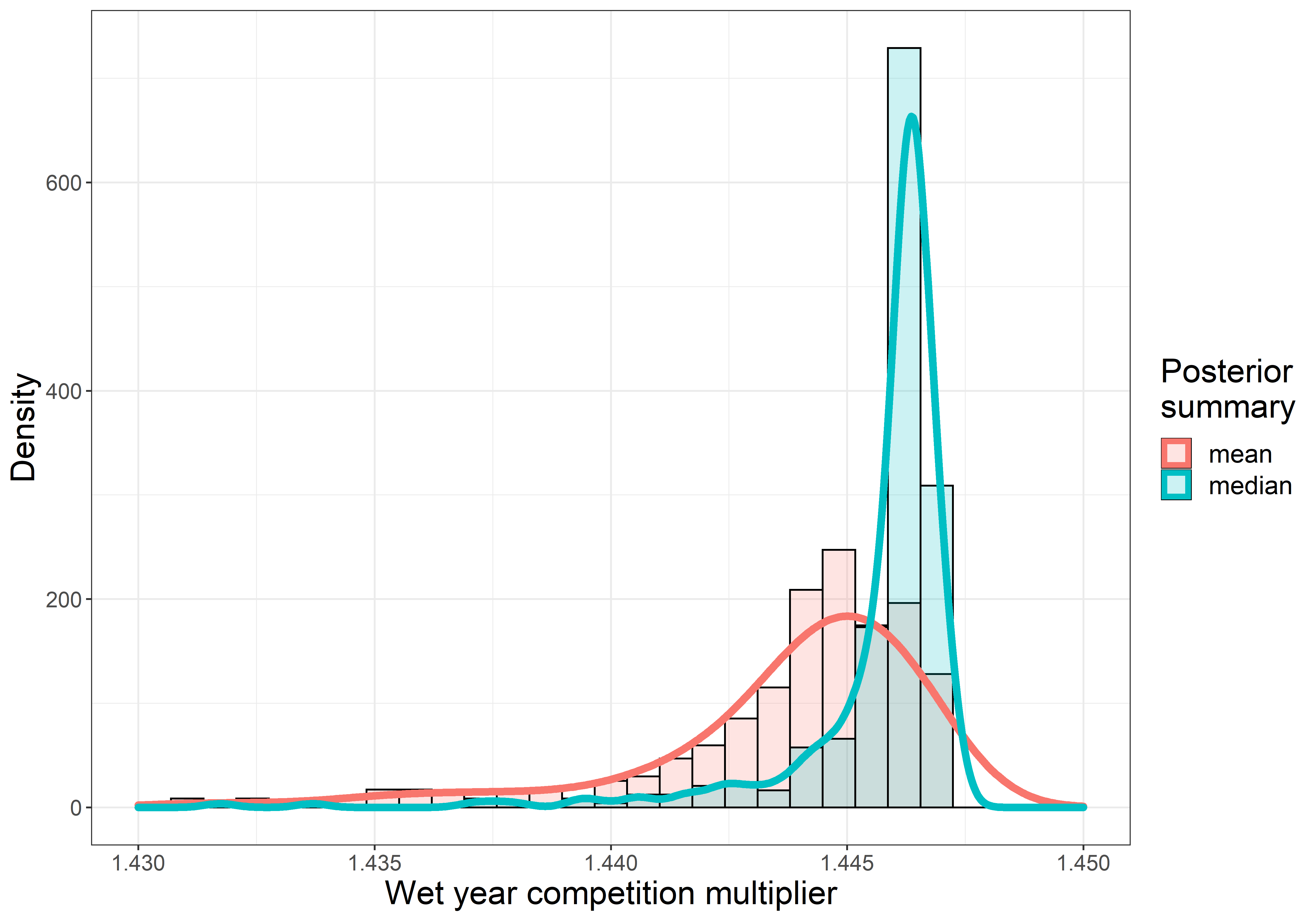}
  \caption{Model-based evidence of ingredient 3: the $EC$ covariance. The x axis shows the factor by which a typically wet year affects the effective competition coefficient. This \textit{Wet year competition multiplier} is defined as $\left(\alpha_{jk} + \gamma_{jk} (\overline{E} + sd(E))\right) / \left(\alpha_{jk} + \gamma_{jk} \overline{E}\right)$. \label{cov}}
\end{figure}

\begin{figure} 
  \centering
      \includegraphics[scale = 0.6]{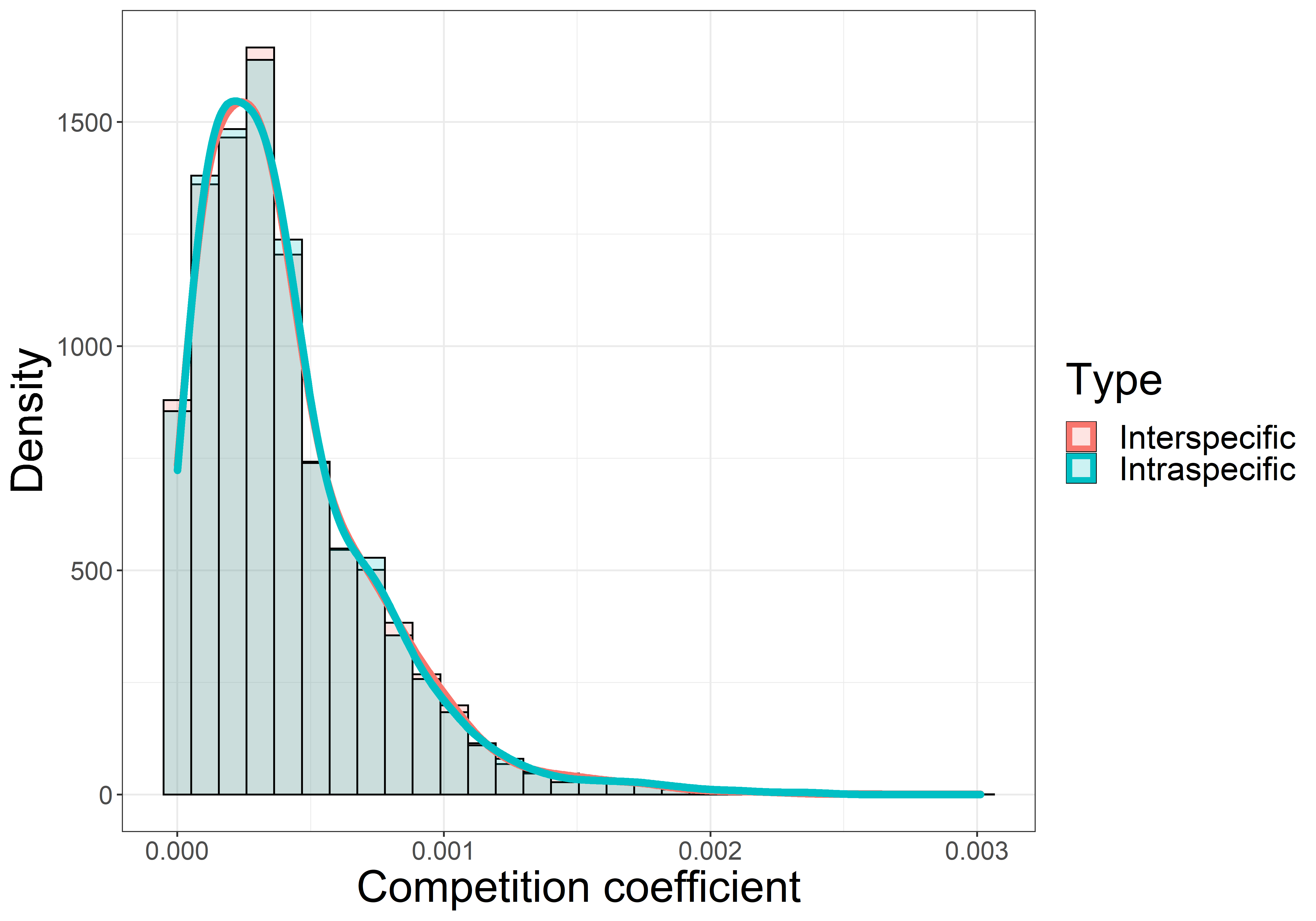}
  \caption{The distribution $19^2 = 361$ competition coefficients. Intra and interspecific coefficients have nearly identical distributions, suggesting that coexistence cannot be attributed to fluctuation-independent mechanisms, e.g., resource partitioning. \label{alpha}}
\end{figure}

\end{appendices}

\newpage

% \section{References}
% \printbibliography[heading=bibintoc]

% \bibliographystyle{apalike}
% \bibliography{refs_for_spatiotemporal_coexistence_paper}

\begin{thebibliography}{}

\bibitem[Abrams, 1984]{abrams1984variability}
Abrams, P. (1984).
\newblock Variability in resource consumption rates and the coexistence of
  competing species.
\newblock {\em Theoretical Population Biology}, 25(1):106--124.

\bibitem[Abrams, 1988]{abrams1988should}
Abrams, P.~A. (1988).
\newblock How should resources be counted?
\newblock {\em Theoretical Population Biology}, 33(2):226--242.

\bibitem[Adler, 2014]{adler2014testing}
Adler, P. (2014).
\newblock Testing the storage effect with long term observational data.
\newblock {\em Temporal dynamics and ecological process}, pages 82--101.

\bibitem[Adler and Drake, 2008]{adler2008environmental}
Adler, P.~B. and Drake, J.~M. (2008).
\newblock Environmental variation, stochastic extinction, and competitive
  coexistence.
\newblock {\em The American Naturalist}, 172(5):E186--E195.

\bibitem[Adler et~al., 2006]{adler2006climate}
Adler, P.~B., HilleRisLambers, J., Kyriakidis, P.~C., Guan, Q., and Levine,
  J.~M. (2006).
\newblock Climate variability has a stabilizing effect on the coexistence of
  prairie grasses.
\newblock {\em Proceedings of the National Academy of Sciences},
  103(34):12793--12798.

\bibitem[Angert et~al., 2009]{angert2009functional}
Angert, A.~L., Huxman, T.~E., Chesson, P., and Venable, D.~L. (2009).
\newblock Functional tradeoffs determine species coexistence via the storage
  effect.
\newblock {\em Proceedings of the National Academy of Sciences},
  106(28):11641--11645.

\bibitem[Armitage and Jones, 2019]{armitage2019negative}
Armitage, D.~W. and Jones, S.~E. (2019).
\newblock Negative frequency-dependent growth underlies the stable coexistence
  of two cosmopolitan aquatic plants.
\newblock {\em Ecology}, 100(5):e02657.

\bibitem[Armitage and Jones, 2020]{armitage2020coexistence}
Armitage, D.~W. and Jones, S.~E. (2020).
\newblock Coexistence barriers confine the poleward range of a globally
  distributed plant.
\newblock {\em Ecology Letters}, 23(12):1838--1848.

\bibitem[Barab{\'a}s et~al., 2018]{barabas2018chesson}
Barab{\'a}s, G., D'Andrea, R., and Stump, S.~M. (2018).
\newblock Chesson's coexistence theory.
\newblock {\em Ecological Monographs}, 88(3):277--303.

\bibitem[Baskin and Baskin, 1998]{baskin1998seeds}
Baskin, C.~C. and Baskin, J.~M. (1998).
\newblock {\em Seeds: ecology, biogeography, and, evolution of dormancy and
  germination}.
\newblock Elsevier.

\bibitem[Bowler et~al., 2022]{bowler2022accounting}
Bowler, C.~H., Weiss-Lehman, C., Towers, I.~R., Mayfield, M.~M., and Shoemaker,
  L.~G. (2022).
\newblock Accounting for demographic uncertainty increases predictions for
  species coexistence: A case study with annual plants.
\newblock {\em Ecology Letters}.

\bibitem[C{\'a}ceres, 1997]{caceres1997temporal}
C{\'a}ceres, C.~E. (1997).
\newblock Temporal variation, dormancy, and coexistence: a field test of the
  storage effect.
\newblock {\em Proceedings of the National Academy of Sciences},
  94(17):9171--9175.

\bibitem[Carpenter et~al., 2017]{carpenter2017stan}
Carpenter, B., Gelman, A., Hoffman, M.~D., Lee, D., Goodrich, B., Betancourt,
  M., Brubaker, M.~A., Guo, J., Li, P., and Riddell, A. (2017).
\newblock Stan: a probabilistic programming language.
\newblock {\em Grantee Submission}, 76(1):1--32.

\bibitem[Carvalho et~al., 2009]{carvalho2009handling}
Carvalho, C.~M., Polson, N.~G., and Scott, J.~G. (2009).
\newblock Handling sparsity via the horseshoe.
\newblock In {\em Artificial Intelligence and Statistics}, pages 73--80. PMLR.

\bibitem[Chesson, 1991]{chesson1991need}
Chesson, P. (1991).
\newblock A need for niches?
\newblock {\em Trends in ecology \& evolution}, 6(1):26--28.

\bibitem[Chesson, 1994]{Chesson1994}
Chesson, P. (1994).
\newblock Multispecies competition in variable environments.
\newblock {\em Theoretical Population Biology}, 45(3):227--276.

\bibitem[Chesson, 2000a]{chesson2000general}
Chesson, P. (2000a).
\newblock General theory of competitive coexistence in spatially-varying
  environments.
\newblock {\em Theoretical Population Biology}, 58(3):211--237.

\bibitem[Chesson, 2000b]{chesson2000mechanisms}
Chesson, P. (2000b).
\newblock Mechanisms of maintenance of species diversity.
\newblock {\em Annual review of Ecology and Systematics}, 31(1):343--366.

\bibitem[Chesson, 2003]{chesson2003quantifying}
Chesson, P. (2003).
\newblock Quantifying and testing coexistence mechanisms arising from
  recruitment fluctuations.
\newblock {\em Theoretical Population Biology}, 64(3):345--357.

\bibitem[Chesson, 2008]{Chesson2008quantifying}
Chesson, P. (2008).
\newblock Quantifying and testing species coexistence mechanisms.
\newblock In {\em Unity in diversity: reflections on ecology after the legacy
  of {Ramon Margalef}}, pages 119--164. Fundacion BBVA Bilbao.

\bibitem[Chesson, 2018]{Chesson2018}
Chesson, P. (2018).
\newblock Updates on mechanisms of maintenance of species diversity.
\newblock {\em Journal of ecology}, 106(5):1773--1794.

\bibitem[Chesson, 2020]{Chesson2019}
Chesson, P. (2020).
\newblock Chesson's coexistence theory: Comment.
\newblock {\em Ecology}, 101(11):e02851.

\bibitem[Chesson et~al., 2004]{Chesson2004}
Chesson, P., Gebauer, R.~L., Schwinning, S., Huntly, N., Wiegand, K., Ernest,
  M.~S., Sher, A., Novoplansky, A., and Weltzin, J.~F. (2004).
\newblock {Resource pulses, species interactions, and diversity maintenance in
  arid and semi-arid environments}.
\newblock {\em Oecologia}, 141(2):236--253.

\bibitem[Chesson and Huntly, 1997]{chesson1997roles}
Chesson, P. and Huntly, N. (1997).
\newblock The roles of harsh and fluctuating conditions in the dynamics of
  ecological communities.
\newblock {\em The American Naturalist}, 150(5):519--553.

\bibitem[Chesson et~al., 2012]{chesson2012storage}
Chesson, P., Huntly, N.~J., Roxburgh, S.~H., Pantastico-Caldas, M., and
  Facelli, J.~M. (2012).
\newblock The storage effect: definition and tests in two plant communities.
\newblock In {\em Temporal dynamics and ecological process}, pages 11--40.
  Cambridge University Press.

\bibitem[Chesson and Kuang, 2010]{chesson2010storage}
Chesson, P. and Kuang, J.~J. (2010).
\newblock The storage effect due to frequency-dependent predation in
  multispecies plant communities.
\newblock {\em Theoretical Population Biology}, 78(2):148--164.

\bibitem[Chesson, 1982]{chesson1982stabilizing}
Chesson, P.~L. (1982).
\newblock The stabilizing effect of a random environment.
\newblock {\em Journal of Mathematical Biology}, 15(1):1--36.

\bibitem[Chesson, 1983]{chesson1983coexistence}
Chesson, P.~L. (1983).
\newblock Coexistence of competitors in a stochastic environment: the storage
  effect.
\newblock In {\em Population biology}, pages 188--198. Springer.

\bibitem[Chesson, 1984]{chesson1984storage}
Chesson, P.~L. (1984).
\newblock The storage effect in stochastic population models.
\newblock In {\em Mathematical ecology}, pages 76--89. Springer.

\bibitem[Chesson, 1985]{chesson1985coexistence}
Chesson, P.~L. (1985).
\newblock Coexistence of competitors in spatially and temporally varying
  environments: a look at the combined effects of different sorts of
  variability.
\newblock {\em Theoretical Population Biology}, 28(3):263--287.

\bibitem[Chesson, 1988]{chesson1988interactions}
Chesson, P.~L. (1988).
\newblock Interactions between environment and competition: how fluctuations
  mediate coexistence and competitive exclusion.
\newblock In {\em Community ecology}, pages 51--71. Springer.

\bibitem[Chesson and Ellner, 1989]{Chesson1989}
Chesson, P.~L. and Ellner, S. (1989).
\newblock {Invasibility and stochastic boundedness in monotonic competition
  models}.
\newblock {\em Journal of Mathematical Biology}, 27(2):117--138.

\bibitem[Chesson and Huntly, 1988]{chesson1988community}
Chesson, P.~L. and Huntly, N. (1988).
\newblock Community consequences of life-history traits in a variable
  environment.
\newblock {\em Annales Zoologici Fennici}, pages 5--16.

\bibitem[Chesson and Warner, 1981]{chesson1981environmentalST}
Chesson, P.~L. and Warner, R.~R. (1981).
\newblock Environmental variability promotes coexistence in lottery competitive
  systems.
\newblock {\em The American Naturalist}, 117(6):923--943.

\bibitem[Cohen, 1966]{cohen1966optimizing}
Cohen, D. (1966).
\newblock Optimizing reproduction in a randomly varying environment.
\newblock {\em Journal of theoretical biology}, 12(1):119--129.

\bibitem[Commoner, 2015]{commoner2015closing}
Commoner, B. (2015).
\newblock The closing circle: nature, man, and technology.
\newblock In {\em Thinking About The Environment}, pages 161--166. Routledge.

\bibitem[Cushing, 1971]{cushing1971dependence}
Cushing, D. (1971).
\newblock The dependence of recruitment on parent stock in different groups of
  fishes.
\newblock {\em ICES Journal of Marine Science}, 33(3):340--362.

\bibitem[Dean and Shnerb, 2020]{dean2020stochasticity}
Dean, A.~M. and Shnerb, N.~M. (2020).
\newblock Stochasticity-induced stabilization in ecology and evolution: a new
  synthesis.
\newblock {\em Ecology}, 101(9):e03098.

\bibitem[DeMalach et~al., 2017]{demalach2017light}
DeMalach, N., Zaady, E., and Kadmon, R. (2017).
\newblock Light asymmetry explains the effect of nutrient enrichment on
  grassland diversity.
\newblock {\em Ecology Letters}, 20(1):60--69.

\bibitem[Dempster, 1955]{Dempster1955}
Dempster, E.~R. (1955).
\newblock {Maintenance of genetic heterogeneity.}
\newblock {\em Cold Spring Harbor symposia on quantitative biology}, 20.

\bibitem[Dennis et~al., 2003]{dennis2003can}
Dennis, B., Desharnais, R.~A., Cushing, J., Henson, S.~M., and Costantino, R.
  (2003).
\newblock Can noise induce chaos?
\newblock {\em Oikos}, 102(2):329--339.

\bibitem[Descamps-Julien and Gonzalez, 2005]{descamps2005stable}
Descamps-Julien, B. and Gonzalez, A. (2005).
\newblock Stable coexistence in a fluctuating environment: an experimental
  demonstration.
\newblock {\em Ecology}, 86(10):2815--2824.

\bibitem[Dormann et~al., 2018]{dormann2018model}
Dormann, C.~F., Calabrese, J.~M., Guillera-Arroita, G., Matechou, E., Bahn, V.,
  Bartoń, K., Beale, C.~M., Ciuti, S., Elith, J., Gerstner, K., Guelat, J.,
  Keil, P., Lahoz-Monfort, J.~J., Pollock, L.~J., Reineking, B., Roberts,
  D.~R., Schröder, B., Thuiller, W., Warton, D.~I., Wintle, B.~A., Wood,
  S.~N., Wüest, R.~O., and Hartig, F. (2018).
\newblock Model averaging in ecology: A review of bayesian,
  information-theoretic, and tactical approaches for predictive inference.
\newblock {\em Ecological Monographs}, 88(4):485--504.

\bibitem[Ellner, 1987]{ellner1987alternate}
Ellner, S. (1987).
\newblock Alternate plant life history strategies and coexistence in randomly
  varying environments.
\newblock {\em Vegetatio}, 69(1):199--208.

\bibitem[Ellner et~al., 2016]{ellner2016quantify}
Ellner, S.~P., Snyder, R.~E., and Adler, P.~B. (2016).
\newblock How to quantify the temporal storage effect using simulations instead
  of math.
\newblock {\em Ecology letters}, 19(11):1333--1342.

\bibitem[Ellner et~al., 2019]{Ellner2019}
Ellner, S.~P., Snyder, R.~E., Adler, P.~B., and Hooker, G. (2019).
\newblock An expanded modern coexistence theory for empirical applications.
\newblock {\em Ecology letters}, 22(1):3--18.

\bibitem[Ellner et~al., 2022]{ellner2022toward}
Ellner, S.~P., Snyder, R.~E., Adler, P.~B., and Hooker, G. (2022).
\newblock Toward a “modern coexistence theory” for the discrete and
  spatial.
\newblock {\em Ecological Monographs}, page e1548.

\bibitem[Facelli et~al., 2005]{facelli2005differences}
Facelli, J.~M., Chesson, P., and Barnes, N. (2005).
\newblock Differences in seed biology of annual plants in arid lands: a key
  ingredient of the storage effect.
\newblock {\em Ecology}, 86(11):2998--3006.

\bibitem[Gause, 1934]{Gause1934}
Gause, G.~F. (1934).
\newblock {\em The struggle for existence.}
\newblock Williams \& Wilkins, Baltimore.

\bibitem[Gelman and Hill, 2007]{gelman2007data}
Gelman, A. and Hill, J. (2007).
\newblock {\em Data analysis using regression and multilevelhierarchical
  models}, volume~1.
\newblock Cambridge University Press New York, NY, USA.

\bibitem[Gillespie, 1977]{gillespie1977natural}
Gillespie, J.~H. (1977).
\newblock Natural selection for variances in offspring numbers: a new
  evolutionary principle.
\newblock {\em The American Naturalist}, 111(981):1010--1014.

\bibitem[Godfrey-Smith, 2009]{godfrey2009darwinian}
Godfrey-Smith, P. (2009).
\newblock {\em Darwinian populations and natural selection}.
\newblock Oxford University Press.

\bibitem[Grainger et~al., 2019]{Grainger2019TheResearch}
Grainger, T.~N., Levine, J.~M., and Gilbert, B. (2019).
\newblock {The Invasion Criterion: A Common Currency for Ecological Research}.
\newblock {\em Trends in Ecology and Evolution}, 34(10):925--935.

\bibitem[Grinell, 1917]{grinell1917niche}
Grinell, J. (1917).
\newblock The niche relationship of california thrsher.
\newblock {\em Auk}, 1:64--82.

\bibitem[Gulisija et~al., 2016]{gulisija2016phenotypic}
Gulisija, D., Kim, Y., and Plotkin, J.~B. (2016).
\newblock Phenotypic plasticity promotes balanced polymorphism in periodic
  environments by a genomic storage effect.
\newblock {\em Genetics}, 202(4):1437--1448.

\bibitem[Gyllenberg and Mesz{\'{e}}na, 2005]{Gyllenberg2005}
Gyllenberg, M. and Mesz{\'{e}}na, G. (2005).
\newblock {On the impossibility of coexistence of infinitely many strategies}.
\newblock {\em Journal of Mathematical Biology}, 50(2):133--160.

\bibitem[Haigh and Smith, 1972]{haigh1972can}
Haigh, J. and Smith, J.~M. (1972).
\newblock Can there be more predators than prey?
\newblock {\em Theoretical Population Biology}, 3(3):290--299.

\bibitem[Haldane and Jayakar, 1963]{Haldane1963}
Haldane, J.~B. and Jayakar, S.~D. (1963).
\newblock {Polymorphism due to selection of varying direction}.
\newblock {\em Journal of Genetics}, 58(2):237--242.

\bibitem[Hallett et~al., 2019]{hallett2019rainfall}
Hallett, L.~M., Shoemaker, L.~G., White, C.~T., and Suding, K.~N. (2019).
\newblock Rainfall variability maintains grass-forb species coexistence.
\newblock {\em Ecology Letters}, 22(10):1658--1667.

\bibitem[Hastie et~al., 2009]{hastie2009elements}
Hastie, T., Tibshirani, R., and Friedman, J. (2009).
\newblock {\em The elements of statistical learning: data mining, inference,
  and prediction}.
\newblock Springer, 2nd ed. edition.

\bibitem[Holt and Chesson, 2014]{holt2014variation}
Holt, G. and Chesson, P. (2014).
\newblock Variation in moisture duration as a driver of coexistence by the
  storage effect in desert annual plants.
\newblock {\em Theoretical Population Biology}, 92:36--50.

\bibitem[Hutchinson, 1959]{hutchinson1959homage}
Hutchinson, G.~E. (1959).
\newblock Homage to santa rosalia or why are there so many kinds of animals?
\newblock {\em The American Naturalist}, 93(870):145--159.

\bibitem[Hutchinson, 1961]{hutchinson1961paradox}
Hutchinson, G.~E. (1961).
\newblock The paradox of the plankton.
\newblock {\em The American Naturalist}, 95(882):137--145.

\bibitem[Ignace et~al., 2018]{ignace2018role}
Ignace, D.~D., Huntly, N., and Chesson, P. (2018).
\newblock The role of climate in the dynamics of annual plants in a chihuahuan
  desert ecosystem.
\newblock {\em Evolutionary Ecology Research}, 19(3):279--297.

\bibitem[Jiang and Morin, 2007]{jiang2007temperature}
Jiang, L. and Morin, P.~J. (2007).
\newblock Temperature fluctuation facilitates coexistence of competing species
  in experimental microbial communities.
\newblock {\em Journal of animal ecology}, 76(4):660--668.

\bibitem[Johnson and Hastings, 2022a]{johnson2022methodsb}
Johnson, E.~C. and Hastings, A. (2022a).
\newblock Methods for calculating coexistence mechanisms: Beyond scaling
  factors.
\newblock {\em Oikos}.

\bibitem[Johnson and Hastings, 2022b]{johnson2022methods}
Johnson, E.~C. and Hastings, A. (2022b).
\newblock Methods for calculating coexistence mechanisms: Beyond scaling
  factors.
\newblock {\em arXiv preprint arXiv:2201.06666}.

\bibitem[Johnson and Hastings, 2022c]{johnson2022towardsb}
Johnson, E.~C. and Hastings, A. (2022c).
\newblock Towards a heuristic understanding of the storage effect.
\newblock {\em Ecology Letters}.

\bibitem[Kelly and Bowler, 2002]{kelly2002coexistence}
Kelly, C.~K. and Bowler, M.~G. (2002).
\newblock Coexistence and relative abundance in forest trees.
\newblock {\em Nature}, 417(6887):437--440.

\bibitem[Kelly and Bowler, 2005]{kelly2005new}
Kelly, C.~K. and Bowler, M.~G. (2005).
\newblock A new application of storage dynamics: differential sensitivity,
  diffuse competition, and temporal niches.
\newblock {\em Ecology}, 86(4):1012--1022.

\bibitem[Klausmeier, 2010]{klausmeier2010successional}
Klausmeier, C.~A. (2010).
\newblock Successional state dynamics: a novel approach to modeling
  nonequilibrium foodweb dynamics.
\newblock {\em Journal of Theoretical Biology}, 262(4):584--595.

\bibitem[Kuang and Chesson, 2009]{Kuang2009CoexistenceEffect}
Kuang, J.~J. and Chesson, P. (2009).
\newblock {Coexistence of annual plants: Generalist seed predation weakens the
  storage effect}.
\newblock {\em Ecology}, 90(1):170--182.

\bibitem[Kuang and Chesson, 2010]{kuang2010interacting}
Kuang, J.~J. and Chesson, P. (2010).
\newblock Interacting coexistence mechanisms in annual plant communities:
  Frequency-dependent predation and the storage effect.
\newblock {\em Theoretical Population Biology}, 77(1):56--70.

\bibitem[Lande, 1998]{lande1998demographic}
Lande, R. (1998).
\newblock Demographic stochasticity and allee effect on a scale with isotropic
  noise.
\newblock {\em Oikos}, pages 353--358.

\bibitem[Letten et~al., 2018]{letten2018species}
Letten, A.~D., Dhami, M.~K., Ke, P.-J., and Fukami, T. (2018).
\newblock Species coexistence through simultaneous fluctuation-dependent
  mechanisms.
\newblock {\em Proceedings of the National Academy of Sciences},
  115(26):6745--6750.

\bibitem[Levin, 1970]{levin1970community}
Levin, S.~A. (1970).
\newblock Community equilibria and stability, and an extension of the
  competitive exclusion principle.
\newblock {\em The American Naturalist}, 104(939):413--423.

\bibitem[Lewontin and Cohen, 1969]{Lewontin1969}
Lewontin, R.~C. and Cohen, D. (1969).
\newblock {On population growth in a randomly varying environment.}
\newblock {\em Proceedings of the National Academy of Sciences of the United
  States of America}, 62(4):1056--1060.

\bibitem[Li and Chesson, 2016]{li2016effects}
Li, L. and Chesson, P. (2016).
\newblock The effects of dynamical rates on species coexistence in a variable
  environment: the paradox of the plankton revisited.
\newblock {\em The American Naturalist}, 188(2):E46--E58.

\bibitem[Lieth and Whittaker, 1973]{lieth1973primary}
Lieth, H. and Whittaker, R.~H. (1973).
\newblock {\em Primary productivity of the biosphere}.
\newblock Springer.

\bibitem[Loreau, 1989]{loreau1989coexistence}
Loreau, M. (1989).
\newblock Coexistence of temporally segregated competitors in a cyclic
  environment.
\newblock {\em Theoretical Population Biology}, 36(2):181--201.

\bibitem[Loreau, 1992]{loreau1992time}
Loreau, M. (1992).
\newblock Time scale of resource dynamics and coexistence through time
  partitioning.
\newblock {\em Theoretical Population Biology}, 41(3):401--412.

\bibitem[Lotka, 1932]{Lotka1932TheSupply}
Lotka, A.~J. (1932).
\newblock {The growth of mixed populations: Two species competing for a common
  food supply}.
\newblock {\em Journal of the Washington Academy of Sciences},
  22(16-17):461--469.

\bibitem[MacArthur, 1967]{MacArthurRobertH1967Ttoi}
MacArthur, R.~H. (1967).
\newblock {\em The theory of island biogeography}.
\newblock Monographs in population biology ; 1. Princeton University Press,
  Princeton, N.J.

\bibitem[May, 1974]{may1974theory}
May, R.~M. (1974).
\newblock On the theory of niche overlap.
\newblock {\em Theoretical population biology}, 5(3):297--332.

\bibitem[Messer et~al., 2016]{messer2016can}
Messer, P.~W., Ellner, S.~P., and Hairston~Jr, N.~G. (2016).
\newblock Can population genetics adapt to rapid evolution?
\newblock {\em Trends in Genetics}, 32(7):408--418.

\bibitem[Mesz{\'e}na et~al., 2006]{meszena2006competitive}
Mesz{\'e}na, G., Gyllenberg, M., P{\'a}sztor, L., and Metz, J.~A. (2006).
\newblock Competitive exclusion and limiting similarity: a unified theory.
\newblock {\em Theoretical Population Biology}, 69(1):68--87.

\bibitem[Metz et~al., 1992]{Metz1992}
Metz, J., Nisbet, R., and Geritz, S. (1992).
\newblock {How should we define ‘fitness’ for general ecological
  scenarios?}
\newblock {\em Trends in Ecology {\&} Evolution}, 7(6):198--202.

\bibitem[Miller and Klausmeier, 2017]{miller2017evolutionary}
Miller, E.~T. and Klausmeier, C.~A. (2017).
\newblock Evolutionary stability of coexistence due to the storage effect in a
  two-season model.
\newblock {\em Theoretical Ecology}, 10(1):91--103.

\bibitem[Pake and Venable, 1995]{pake1995coexistence}
Pake, C.~E. and Venable, D.~L. (1995).
\newblock Is coexistence of sonoran desert annuals mediated by temporal
  variability reproductive success.
\newblock {\em Ecology}, 76(1):246--261.

\bibitem[Pake and Venable, 1996]{pake1996seed}
Pake, C.~E. and Venable, D.~L. (1996).
\newblock Seed banks in desert annuals: implications for persistence and
  coexistence in variable environments.
\newblock {\em Ecology}, 77(5):1427--1435.

\bibitem[Pande et~al., 2020]{pande2020mean}
Pande, J., Fung, T., Chisholm, R., and Shnerb, N.~M. (2020).
\newblock Mean growth rate when rare is not a reliable metric for persistence
  of species.
\newblock {\em Ecology letters}, 23(2):274--282.

\bibitem[Petry et~al., 2018]{petry2018competition}
Petry, W.~K., Kandlikar, G.~S., Kraft, N.~J., Godoy, O., and Levine, J.~M.
  (2018).
\newblock A competition--defence trade-off both promotes and weakens
  coexistence in an annual plant community.
\newblock {\em Journal of Ecology}, 106(5):1806--1818.

\bibitem[{R Core Team}, 2022]{rcore2022}
{R Core Team} (2022).
\newblock {\em R: A Language and Environment for Statistical Computing}.
\newblock R Foundation for Statistical Computing, Vienna, Austria.

\bibitem[Rees, 1994]{rees1994delayed}
Rees, M. (1994).
\newblock Delayed germination of seeds: a look at the effects of adult
  longevity, the timing of reproduction, and population age/stage structure.
\newblock {\em The American Naturalist}, 144(1):43--64.

\bibitem[Rees, 2013]{rees2013competition}
Rees, M. (2013).
\newblock Competition on productivity gradients--what do we expect?
\newblock {\em Ecology Letters}, 16(3):291--298.

\bibitem[Reinhold, 2000]{reinhold2000maintenance}
Reinhold, K. (2000).
\newblock Maintenance of a genetic polymorphism by fluctuating selection on
  sex-limited traits.
\newblock {\em Journal of Evolutionary Biology}, 13(6):1009--1014.

\bibitem[Rosenzweig, 1968]{rosenzweig1968net}
Rosenzweig, M.~L. (1968).
\newblock Net primary productivity of terrestrial communities: prediction from
  climatological data.
\newblock {\em The American Naturalist}, 102(923):67--74.

\bibitem[Sala et~al., 1988]{sala1988primary}
Sala, O.~E., Parton, W.~J., Joyce, L., and Lauenroth, W. (1988).
\newblock Primary production of the central grassland region of the united
  states.
\newblock {\em Ecology}, 69(1):40--45.

\bibitem[Sale, 1977]{Sale1977}
Sale, P.~F. (1977).
\newblock {Maintenance of High Diversity in Coral Reef Fish Communities}.
\newblock {\em The American Naturalist}, 111(978):337--359.

\bibitem[Schreiber, 2021]{schreiber2021positively}
Schreiber, S.~J. (2021).
\newblock Positively and negatively autocorrelated environmental fluctuations
  have opposing effects on species coexistence.
\newblock {\em The American Naturalist}, 197(4):000--000.

\bibitem[Schreiber et~al., 2019]{schreiber2019rarity}
Schreiber, S.~J., Yamamichi, M., and Strauss, S.~Y. (2019).
\newblock When rarity has costs: coexistence under positive
  frequency-dependence and environmental stochasticity.
\newblock {\em Ecology}, 100(7):e02664.

\bibitem[Sears and Chesson, 2007]{sears2007new}
Sears, A.~L. and Chesson, P. (2007).
\newblock New methods for quantifying the spatial storage effect: an
  illustration with desert annuals.
\newblock {\em Ecology}, 88(9):2240--2247.

\bibitem[Snyder, 2012]{snyder2012storage}
Snyder, R. (2012).
\newblock Storage effect.
\newblock In {\em Encyclopedia of theoretical ecology}, pages 722--726.
  University of California Press, 1st ed. edition.

\bibitem[Stearns, 2000]{Stearns2000}
Stearns, S.~C. (2000).
\newblock Daniel bernoulli (1738): evolution and economics under risk.
\newblock {\em Journal of biosciences}, 25(3):221--228.

\bibitem[Stump and Chesson, 2017]{stump2017optimally}
Stump, S.~M. and Chesson, P. (2017).
\newblock How optimally foraging predators promote prey coexistence in a
  variable environment.
\newblock {\em Theoretical Population Biology}, 114:40--58.

\bibitem[Szuwalski et~al., 2015]{szuwalski2015examining}
Szuwalski, C.~S., Vert-Pre, K.~A., Punt, A.~E., Branch, T.~A., and Hilborn, R.
  (2015).
\newblock Examining common assumptions about recruitment: a meta-analysis of
  recruitment dynamics for worldwide marine fisheries.
\newblock {\em Fish and Fisheries}, 16(4):633--648.

\bibitem[Towers et~al., 2020]{towers2020requirements}
Towers, I.~R., Bowler, C.~H., Mayfield, M.~M., and Dwyer, J.~M. (2020).
\newblock Requirements for the spatial storage effect are weakly evident for
  common species in natural annual plant assemblages.
\newblock {\em Ecology}, 101(12):e03185.

\bibitem[Turelli, 1978]{turelli1978reexamination}
Turelli, M. (1978).
\newblock A reexamination of stability in randomly varying versus deterministic
  environments with comments on the stochastic theory of limiting similarity.
\newblock {\em Theoretical Population Biology}, 13(2):244--267.

\bibitem[Usinowicz et~al., 2017]{usinowicz2017temporal}
Usinowicz, J., Chang-Yang, C.-H., Chen, Y.-Y., Clark, J.~S., Fletcher, C.,
  Garwood, N.~C., Hao, Z., Johnstone, J., Lin, Y., Metz, M.~R., Masaki, T.,
  Nakashizuka, T., Sun, I.-F., Valencia, R., Wang, Y., Zimmerman, J.~K., Ives,
  A.~R., and Wright, S.~J. (2017).
\newblock Temporal coexistence mechanisms contribute to the latitudinal
  gradient in forest diversity.
\newblock {\em Nature}, 550(7674):105--108.

\bibitem[Usinowicz et~al., 2012]{usinowicz2012coexistence}
Usinowicz, J., Wright, S.~J., and Ives, A.~R. (2012).
\newblock Coexistence in tropical forests through asynchronous variation in
  annual seed production.
\newblock {\em Ecology}, 93(9):2073--2084.

\bibitem[Vehtari et~al., 2019]{vehtari2019loo}
Vehtari, A., Gabry, J., Magnusson, M., Yao, Y., and Gelman, A. (2019).
\newblock loo: Efficient leave-one-out cross-validation and waic for bayesian
  models.
\newblock R package version 2.2.0.

\bibitem[Venable, 2007]{venable2007bet}
Venable, D.~L. (2007).
\newblock Bet hedging in a guild of desert annuals.
\newblock {\em Ecology}, 88(5):1086--1090.

\bibitem[Venable et~al., 1993]{venable1993diversity}
Venable, D.~L., Pake, C.~E., and Caprio, A.~C. (1993).
\newblock Diversity and coexistence of sonoran desert winter annuals.
\newblock {\em Plant Species Biology}, 8(2-3):207--216.

\bibitem[Volterra, 1926]{volterra1926variationsST}
Volterra, V. (1926).
\newblock Variations and fluctuations of the number of individuals in animal
  species living together.
\newblock {\em Animal Ecology}, pages 409--448.

\bibitem[Warner and Chesson, 1985]{warner1985coexistence}
Warner, R.~R. and Chesson, P.~L. (1985).
\newblock Coexistence mediated by recruitment fluctuations: a field guide to
  the storage effect.
\newblock {\em The American Naturalist}, 125(6):769--787.

\bibitem[Warner and Hoffman, 1980]{warner1980population}
Warner, R.~R. and Hoffman, S.~G. (1980).
\newblock Population density and the economics of territorial defense in a
  coral reef fish.
\newblock {\em Ecology}, 61(4):772--780.

\bibitem[Weiss-Lehman et~al., 2022]{weiss2022disentangling}
Weiss-Lehman, C.~P., Werner, C.~M., Bowler, C.~H., Hallett, L.~M., Mayfield,
  M.~M., Godoy, O., Aoyama, L., Barab{\'a}s, G., Chu, C., Ladouceur, E., et~al.
  (2022).
\newblock Disentangling key species interactions in diverse and heterogeneous
  communities: A bayesian sparse modelling approach.
\newblock {\em Ecology Letters}, 25(5):1263--1276.

\bibitem[Wittgenstein, 1968]{wittgenstein1968philosophical}
Wittgenstein, L. (1968).
\newblock {\em Philosophical Investigations}.
\newblock Basil Blackwell.

\bibitem[Yamamichi and Hoso, 2017]{yamamichi2017roles}
Yamamichi, M. and Hoso, M. (2017).
\newblock Roles of maternal effects in maintaining genetic variation: maternal
  storage effect.
\newblock {\em Evolution}, 71(2):449--457.

\bibitem[Yuan and Chesson, 2015]{Yuan2015}
Yuan, C. and Chesson, P. (2015).
\newblock The relative importance of relative nonlinearity and the storage
  effect in the lottery model.
\newblock {\em Theoretical population biology}, 105:39--52.

\bibitem[Zepeda and Martorell, 2019]{zepeda2019fluctuation}
Zepeda, V. and Martorell, C. (2019).
\newblock Fluctuation-independent niche differentiation and relative
  non-linearity drive coexistence in a species-rich grassland.
\newblock {\em Ecology}, 100(8):e02726.

\end{thebibliography}

\end{document}